\documentclass[journal]{IEEEtran}

\usepackage{mathrsfs}
\usepackage[noadjust]{cite}
\usepackage{graphicx,color,overpic,psfrag}
\usepackage{amsmath, amssymb}
\usepackage{latexsym}
\usepackage{bm}
\usepackage{amssymb}
\usepackage{cases}
\usepackage{array}
\usepackage{fancyhdr}
\usepackage{setspace}

\ifCLASSOPTIONcompsoc
\usepackage[caption=false,font=normalsize,labelfont=sf,textfont=sf]{subfig}
\else
\usepackage[caption=false,font=footnotesize]{subfig}
\fi

\usepackage{url}
\usepackage{algpseudocode}
\usepackage{algorithm}
\usepackage{blkarray}
\usepackage{booktabs}

\usepackage{multirow}
\usepackage{dsfont}
\usepackage{tabularx}
\usepackage[table]{xcolor}

\usepackage{amsfonts}

\usepackage{letltxmacro}

\graphicspath{{figure/}}%path of the figures

\newcommand{\ra}[1]{\renewcommand{\arraystretch}{#1}}

\newtheorem{remark}{Remark}

\newtheorem{prop}{Proposition}

\newcommand{\figref}[1]{Fig. \ref{#1}}
\newcommand{\tabref}[1]{Table \ref{#1}}
\newcommand{\alref}[1]{\textbf{Algorithm \ref{#1}}}

\newcommand{\secref}[1]{Section \ref{#1}}

\newcommand{\propref}[1]{\emph{Proposition \ref{#1}}}
\newcommand{\Exp}{{\mathsf{E}}}
\newcommand{\expect}[1]{\Exp\left\{#1\right\}}
\newcommand{\tr}[1]{\mathsf{tr}\left\{#1\right\}}
\newcommand{\diag}[1]{\mathsf{diag}\left\{#1\right\}}

\newcommand{\cF}{\mathcal{F}}

\newcommand{\cK}{\mathcal{K}}

\newcommand{\cN}{\mathcal{N}}
\newcommand{\cO}{\mathcal{O}}
\newcommand{\cP}{\mathcal{P}}

\newcommand{\cS}{\mathcal{S}}

\newcommand{\ba}{\mathbf{a}}

\newcommand{\bc}{\mathbf{c}}

\newcommand{\bn}{\mathbf{n}}

\newcommand{\bs}{\mathbf{s}}

\newcommand{\bu}{\mathbf{u}}

\newcommand{\bx}{\mathbf{x}}
\newcommand{\by}{\mathbf{y}}
\newcommand{\bz}{\mathbf{z}}

\newcommand{\bA}{\mathbf{A}}
\newcommand{\bB}{\mathbf{B}}
\newcommand{\bC}{\mathbf{C}}
\newcommand{\bD}{\mathbf{D}}
\newcommand{\bE}{\mathbf{E}}

\newcommand{\bG}{\mathbf{G}}
\newcommand{\bH}{\mathbf{H}}
\newcommand{\bI}{\mathbf{I}}

\newcommand{\bQ}{\mathbf{Q}}

\newcommand{\bU}{\mathbf{U}}
\newcommand{\bV}{\mathbf{V}}
\newcommand{\bW}{\mathbf{W}}

\newcommand{\bzero}{\mathbf{0}}

\newcommand{\bLambda}{{\boldsymbol\Lambda}}

\newcommand{\bGamma}{{\boldsymbol\Gamma}}
\newcommand{\bgamma}{{\boldsymbol\gamma}}
\newcommand{\bPsi}{{\boldsymbol\Psi}}
\newcommand{\bpsi}{{\boldsymbol\psi}}
\newcommand{\bPhi}{{\boldsymbol\Phi}}
\newcommand{\bphi}{{\boldsymbol\phi}}
\newcommand{\bOmega}{{\boldsymbol\Omega}}

\newcommand{\ntb}{\notag\\}

\newcommand{\R}{\mathbb{R}}
\newcommand{\C}{\mathbb{C}}

\newcommand{\rmG}{\mathrm{G}}
\newcommand{\rmR}{\mathrm{R}}

\newcommand{\SE}{\rho_{\mathrm{SE}}}
\newcommand{\EE}{\rho_{\mathrm{EE}}}
\newcommand{\RE}{\rho_{\mathrm{RE}}}
\newcommand{\SEt}{{\widetilde{\rho}}_{\mathrm{SE}}}
\newcommand{\EEt}{{\widetilde{\rho}}_{\mathrm{EE}}}

\newcommand{\SEde}{{\overline{\rho}}_{\mathrm{SE}}}

\newcommand{\Pck}{P_{\mathrm{c},k}}
\newcommand{\Ps}{P_{\mathrm{s}}}

\newcommand{\Pmaxk}{P_{\mathrm{max},k}}
\newcommand{\Pmax}{P_{\mathrm{max}}}
\newcommand{\s}{\frac{1}{\sigma ^{2}}}

\newcommand{\tot}{\mathrm{tot}}
\newcommand{\Ptot}{P_{\mathrm{tot}}}
\newcommand{\Psum}{P_{\mathrm{sum}}}

\newcommand{\beamH}{\widetilde{\bH}}

\newcommand{\Real}[1]{\Re \left\{#1\right\}}
\newcommand{\Imag}[1]{\Im \left\{#1\right\}}
\allowdisplaybreaks

\begin{document}

\title{Energy Efficiency and Spectral Efficiency Tradeoff in RIS-Aided Multiuser MIMO Uplink Transmission}

\author{Li~You, Jiayuan~Xiong, Derrick~Wing~Kwan~Ng, Chau~Yuen, Wenjin~Wang, and~Xiqi~Gao
%Li~You,~\IEEEmembership{Member,~IEEE}, Jiayuan~Xiong,~\IEEEmembership{Student~Member,~IEEE}, Derrick~Wing~Kwan~Ng,~\IEEEmembership{Senior~Member,~IEEE}, Chau~Yuen,~\IEEEmembership{Senior Member,~IEEE}, Wenjin~Wang,~\IEEEmembership{Member,~IEEE}, and~Xiqi~Gao,~\IEEEmembership{Fellow,~IEEE}
%%\thanks{Manuscript received November 11, 2016; revised February 20, 2017; accepted March 6, 2017.}%
%%\thanks{This work was supported by National Natural Science Foundation of China under Grants 61320106003, 61471113, 61521061 and 61631018, the China High-Tech 863 Plan under Grants 2015AA01A701 and 2014AA01A704, National Science and Technology Major Project of China under Grant 2014ZX03003006-003, and the Huawei Cooperation Project. Part of this work has been submitted to IEEE ICC'17.
%%}% <-this % stops a space
\thanks{Part of this work will be presented at the 2020 IEEE Global Communications Conference (GLOBECOM) \cite{xiong2020Globecom}.
}% <-this % stops a space
\thanks{
Li You, Jiayuan Xiong, Wenjin Wang, and Xiqi Gao are with the National Mobile Communications Research Laboratory, Southeast University, Nanjing 210096, China, and also with the Purple
Mountain Laboratories, Nanjing 211100, China (e-mail: liyou@seu.edu.cn; jyxiong@seu.edu.cn; wangwj@seu.edu.cn; xqgao@seu.edu.cn).
}% <-this % stops a space
\thanks{
Derrick Wing Kwan Ng is with the School of Electrical Engineering and Telecommunications, University of New South Wales, Sydney, NSW 2052, Australia (e-mail: w.k.ng@unsw.edu.au).
}% <-this % stops a space
\thanks{
Chau Yuen is with the Singapore University of Technology and Design (SUTD), Singapore 487372 (e-mail: yuenchau@sutd.edu.sg).
}% <-this % stops a space
}
\maketitle
\begin{abstract}
  The emergence of reconfigurable intelligent surfaces (RISs) enables us to establish programmable radio wave propagation that caters for wireless communications, via employing low-cost passive reflecting units. This work studies the non-trivial tradeoff between energy efficiency (EE) and spectral efficiency (SE) in multiuser multiple-input multiple-output (MIMO) uplink communications aided by a RIS equipped with discrete phase shifters. For reducing the required signaling overhead and energy consumption, our transmission strategy design is based on the partial channel state information (CSI), including the statistical CSI between the RIS and user terminals (UTs) and the instantaneous CSI between the RIS and the base station. To investigate the EE-SE tradeoff, we develop a framework for the joint optimization of UTs' transmit precoding and RIS reflective beamforming to maximize a performance metric called resource efficiency (RE). For the design of UT's precoding, it is simplified into the design of UTs' transmit powers with the aid of the closed-form solutions of UTs' optimal transmit directions. To avoid the high complexity in computing the nested integrals involved in the expectations, we derive an asymptotic deterministic objective expression. For the design of the RIS phases, an iterative mean-square error minimization approach is proposed via capitalizing on the homotopy, accelerated projected gradient, and majorization-minimization methods. Numerical results illustrate the effectiveness and rapid convergence rate of our proposed optimization framework.
\end{abstract}


\section{Introduction}
Recently, an emerging concept called reconfigurable intelligent surface (RIS)\footnote{In the literature, RIS with different variations is also referred to as several other terms, e.g., intelligent reflecting surface (IRS), large intelligent surface (LIS), digitally controllable scatterer (DCS), and software controllable surface (SCS), etc.} has been proposed and quickly gained tremendous research attentions \cite{qingqing2019towards}. Generally, RISs are artificial metamaterial structures composed of adaptive composite material layers that can reflect incident electromagnetic waves to specific directions via applying external stimuli \cite{liaskos2018new}. Specifically, the composite material sheets are constituted by a huge number of dielectric patches, which are usually constructed by low-cost passive scattering/reflection elements. One of the major differences between a RIS and a conventional surface, e.g., a wall, lies in the surface electric currents induced by an impinging radio wave, which can control how the surface reacts to the impinging radio wave \cite{di2019smart,huang2019holographic}. For a conventional wall, the reflection characteristics are fixed by nature, i.e., governed by the physical laws, and therefore can not be artificially changed. In contrast, for a RIS, the incident and reflection angles can be designed since the reflection properties of the RIS scattering elements can be intelligently adjusted through some integrated electronics \cite{basar2019wireless}. Due to the great adaptability to time-varying wireless propagation environments, RISs are deemed as a promising enabler for realizing controllable and reconfigurable propagation environments, referred to as smart radio environments \cite{huang2019holographic}. More precisely, RISs are capable of customizing wireless environments via shaping the reflection of the impinging radio waves with adjustable phase shifts to satisfy certain system requirements. In addition, RISs can be conveniently installed on or removed from, e.g., ceilings of buildings and walls, with a quite low-implementation cost due to the relatively low hardware footprints \cite{wu2019intelligent,huang2019reconfigurable,zhou2020,huang2020reconfigurable}. These appealing features position RIS an energy-saving and cost-effective technology for providing superior system performance, which is progressively entering the mainstream of communication networks \cite{wu2019beamforming,ElMossallamy20Reconfigurable,Zhang2019Multiple,Renzo2020smart}.

Owing to the enormous potential benefits promised by the RIS-empowered environments, various research activities have been recently devoted in the literature, covering different aspects, e.g., channel modeling \cite{ozdogan2019intelligent,di2020analytical}, channel estimation \cite{lin2019channel,you2019progressive}, modulation and encoding \cite{karasik2019beyond,basar2020reconfigurable}, and performance evaluation in RIS-aided wireless networks \cite{jung2020performance,Gao20Reconfigurable,Zou20Joint}. In this work, we focus on the design of resource allocation in RIS-aided wireless transmissions. In this connection, a large body of works was committed to the enhancement of the system spectral efficiency (SE). For instance, maximizing the achievable rate of a single-user multiple-input-single-output (MISO) link was studied in \cite{abeywickrama2020intelligent}. In particular, a practical modeling of the RIS units was considered where the phases of the reflection coefficients depend on the amplitudes. In \cite{xie2020max}, the authors investigated the inter-cell interference suppression in a RIS-aided multi-cell MISO system. In addition to SE, energy efficiency (EE) is another important performance metric in the research contributions on RIS-aided wireless communications \cite{you2020reconfigurable,huang2019reconfigurable}. For example, in \cite{huang2019reconfigurable}, an investigation on the benefits of applying RISs for green communications was carried out, where the phases of a RIS and downlink transmit powers were jointly designed to maximize the EE of the RIS-aided downlink multi-user MISO system. However, in some cases, the maximization of EE and SE can not coincide \cite{tang2014resource,mahapatra2015energy}. In particular, the maximum EE is sometimes achieved at the price of SE degradation, and increasing SE sometimes leads to a loss of EE. Hence, the topic of how to tradeoff between EE and SE is also worthy of investigation. To our best knowledge, this is the first paper providing an optimization framework to achieve the EE-SE tradeoff in RIS-aided multi-user multiple-input multiple-output (MIMO) uplink networks.

The performance of RIS-empowered systems highly depends on the adjustment of the passive elements at the RIS. How to properly and effectively adapt the reflection coefficient of each RIS unit according to the channel state information (CSI) is one of the major engineering challenges in the design of RIS-aided transmissions. In this context, a plethora of investigations were based on the assumption that the passive reflecting elements equipped at RISs can be implemented and modeled via continuous phase shifters \cite{wu2019intelligent,huang2019reconfigurable,zhao2019intelligent,wang2020Intelligent,yu2019miso,you2020reconfigurable}. However, it is in fact expensive to realize infinite-resolution phase shifters due to limitations of hardware implementations \cite{guo2020Weighted}. Moreover, the energy consumption of each phase shifter increases with its bit resolution. Hence, it is more practicable to implement the RIS reflection units via adapting phase shifters with finite resolutions \cite{wu2019beamforming,cui2019hybrid,di2020practical,omid2020irs,Zhang20Reconfigurable}. Another critical design issue is how often one should adjust the phase shifts of the RIS in practice. Although the real-time adjustment of the RIS phases based on full instantaneous CSI is more preferred in establishing a favorable communication environment, it is challenging to be realized in mobility scenarios due to, e.g., exceedingly high required energy consumption and significant signaling overhead \cite{he2017geometrical,he2019propagation,Zappone2020Overhead}. In practice, the RIS phase tuning is performed by adapting the biasing voltages through a smart controller, which usually dominates the energy consumption at a RIS \cite{wu2019beamforming}. Therefore, it is not energy-efficient to update the RIS phases too frequently, especially adapting to the fast time-varying instantaneous CSI. As such, considering the fast-moving user terminals (UTs) and the corresponding fast time-varying UT-to-RIS channels, we exploit the more slow time-varying channel characteristics, i.e., the statistical CSI, to develop practical resource allocation strategies in the considered RIS-aided uplink communications.

Given these motivations, we study the transmission strategies for RIS-aided multi-user MIMO uplink systems with the assumption of partial CSI, taking into account both continuous and discrete RIS phase shift values. Considering both EE and SE as the system design criteria, we adopt a performance measure called resource efficiency (RE), which is suitable and flexible to strike a balance between EE and SE \cite{tang2014resource}. To address the RE maximization problem, we then develop an optimization framework for the joint design of the transmit beamforming at the UT sides and the phase shift values at the RIS. The main contributions of this paper are summarized as follows:
\begin{itemize}
\item We investigate the RE maximization transmission design to attain an EE-SE tradeoff in the RIS-aided multi-user MIMO uplink systems. For the RIS, we consider two assumptions of phase shifters with either continuous or discrete values. For the knowledge of channels, we focus on a practical scenario with partial CSI, involving the instantaneous CSI of the slowly time-varying channel from the fixed RIS to the fixed base station (BS) as well as the statistical CSI of the fast time-varying UT-to-RIS channels.
\item We develop an optimization framework via leveraging the alternating optimization (AO) method to iteratively update the UTs' transmit covariance matrices and the RIS phase shift values. For the UTs' transmit covariance matrices, we begin with deriving closed-form optimal solutions for characterizing UTs' transmit signal directions. Then, to reduce the complexity in computing the nested integrals in the objective function, a simple and asymptotic SE expression is proposed. Accordingly, we utilize the quadratic transformation to acquire asymptotically suboptimal solutions for UTs' power allocation matrices.
\item We address the optimization of the RIS phase shift values by handling an equivalent mean-squared error (MSE) minimization problem, which is further converted into a convex constrained problem via using the homotopy optimization method. Then, we address it by means of an inexact majorization-minimization (MM) method. The proposed approach is applicable to both settings with continuous and discrete phase shift values.
\item Uniting all these techniques adopted above forms the overall optimization framework for the RE maximization (as well as the EE or SE maximization) problem in the RIS-empowered multi-user MIMO uplink system. Despite the consideration of partial CSI, the proposed framework can also be applicable to the case of instantaneous CSI with slight modification. Simulation results are conducted to demonstrate the potential performance gains reaped by our proposed framework against various baseline schemes.
\end{itemize}

The rest of this paper is organized as follows. In \secref{sec:sysmod}, we outline the system model of the RIS-aided multi-user MIMO uplink system and then introduce the problem formulation. In \secref{sec:optimization}, we develop an algorithm framework for the RE maximization problem, including the optimization approaches for the transmit covariance matrices at the UT sides and the phase shifts adopted at the RIS. In \secref{sec:numerical}, simulation results are presented to numerically analyze the developed optimization framework. Finally, \secref{sec:conclusion} provides the concluding remarks. Moreover, we list the adopted notations throughout this paper in \tabref{tab:notations} for clarity.

\newcolumntype{L}{>{\hspace*{-\tabcolsep}}l}
 \newcolumntype{R}{c<{\hspace*{-\tabcolsep}}}
 \definecolor{lightblue}{rgb}{0.93,0.95,1.0}
 \begin{table}[!t]\label{tab:notations}
  \caption{Notation List}
  \centering
\footnotesize
  \ra{1.2}
 \begin{tabular}{LR}
  \toprule
  Notation &   {Definition}\\
  \midrule
  \rowcolor{lightblue}
  $\triangleq$ & {Definition} \\
  $\bA$ & {Matrix} \\
  \rowcolor{lightblue}
  $(\cdot)^{H}$ & {Conjugate transpose} \\
  $\expect{ \cdot }$ & {Expectation} \\
  \rowcolor{lightblue}
  $\odot$ & {Hadamard product} \\
  $[\bA]_{m,n}$ & {The $(m,n)$th entry of $\bA$} \\
  \rowcolor{lightblue}
  $\bx$ & {Column vector} \\
  $\bzero$ & {Zero matrix/vector} \\
  \rowcolor{lightblue}
  $\mathcal{CN}(\ba,\bB)$ & {Circular symmetric complex Gaussian distribution} \\
  $\bI_M$ & {$M \times M$ identity matrix} \\
  \rowcolor{lightblue}
  $\diag{\cdot}$ & {Diagonalization operator} \\
  $\jmath = \sqrt{-1}$ & {Imaginary unit} \\
  \rowcolor{lightblue}
  $\det(\bA)$ & {Determinant of $\bA$} \\
  $\tr{\cdot}$ & {Matrix trace} \\
  \rowcolor{lightblue}
  $(\cdot)^{-1}$ & {Matrix inverse} \\
  $\bA \succeq \bzero$ & {Positive semi-definite matrix} \\
  \rowcolor{lightblue}
  $\infty$ & {Infinity} \\
  $(\cdot)^{T}$ & {Transpose} \\
  \rowcolor{lightblue}
  $\| \bx \|$ & {Euclidean norm of a vector $\bx$} \\
  $\Real{\cdot}$ & {Real part of the input} \\
  \rowcolor{lightblue}
  $\Imag{\cdot}$ & {Imaginary part of the input} \\
  $\nabla_\bx f(\bx)$ & {Gradient of $f$ at $\bx$} \\
  \rowcolor{lightblue}
  $\langle \bx,\by \rangle = \Real{\bx^H \by}$ & {Inner product} \\
  $\angle \phi$ & {Argument of a complex number $\phi$} \\
  \rowcolor{lightblue}
  $\mathcal{O}\left(\cdot\right)$ & {Computational complexity} \\
  \bottomrule
 \end{tabular}
 \end{table}

\section{System Model}\label{sec:sysmod}

\begin{figure}[!t]
\centering
\includegraphics[width=0.45\textwidth]{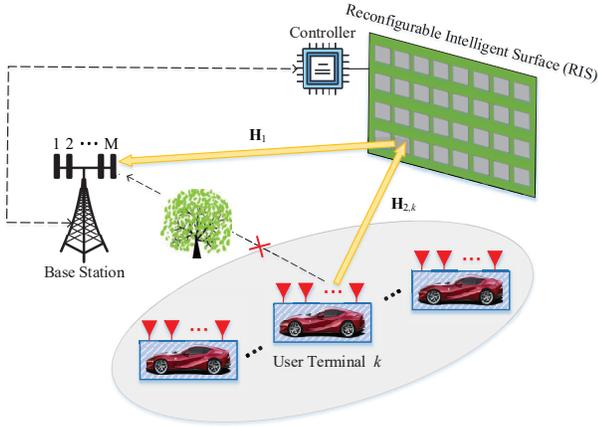}
\caption{The considered RIS-aided multi-user MIMO uplink system.}
\label{fig:RIS}
\end{figure}

For ease of exposition, the considered RIS-aided multi-user MIMO transmission is illustrated in \figref{fig:RIS}, where a total of $K$ multiple-antenna UTs send messages simultaneously to an $M$-antenna BS. We denote $\cK \triangleq \left\{ {1,2, \ldots ,K} \right\}$ as the UT set and the number of UT antennas is assumed to be $N_k$ at UT $k\in\cK$. The communication from all the UTs to the BS is aided by a RIS with a set of $N_{\rmR}$ reflecting elements, which is denoted by $\cN \triangleq \left\{ {1,2, \ldots ,N_{\rmR}} \right\}$. Owing to a smart controller installed at the RIS, all the reflecting units are programmable to control the reflection reacting to the incident signals in real-time \cite{you2019progressive,abeywickrama2020intelligent}. It is assumed that the direct UT-to-BS transmissions are blocked due to unfavorable propagation conditions, e.g., substantial obstructions, thus being neglected in the system model, as commonly adopted in e.g., \cite{you2019progressive,huang2019reconfigurable}. In addition, only the first-time-reflected signals are considered while the reflections more than once with negligible power are ignored attribute to e.g., substantially high path loss \cite{pan2019multicell,abeywickrama2020intelligent}.

\subsection{System Model}
We denote the channel matrix of the RIS-to-BS link as $\bH_1 \in \C^{M \times N_{\rmR}}$, and the channel from UT $k$ to the RIS as $\bH_{2,k} \in \C^{N_{\rmR} \times N_k},\forall k$. The jointly correlated Rayleigh fading model \cite{Gao09Statistical} is applied to describe the UT-to-RIS channel spatial correlations. Mathematically, $\bH_{2,k}$ takes the form as follows:
\begin{align}\label{eq:beam_H2}
\bH_{2,k} & = \bU_{2,k} \beamH_{2,k} \bV^H_{2,k},\quad \forall k \in \cK,
\end{align}
where the unitary matrices $\bU_{2,k} \in \C^{N_{\rmR} \times N_{\rmR}}$ and $\bV_{2,k} \in \C^{N_k \times N_k}$ are both deterministic. Note that $\bU_{2,k}$ and $\bV_{2,k}$ represent the eigenvector matrices of the receive and transmit correlation matrices of $\bH_{2,k}$, respectively \cite{Gao09Statistical}. In addition, the complex-valued matrix $\beamH_{2,k}$ is random and all entries of $\beamH_{2,k}$ are zero mean and independently Gaussian distributed.

As stated previously, we consider a scenario with partial CSI, in which the instantaneous knowledge of the RIS-to-BS channel $\bH_1$ can be perfectly known by adopting some existing methods such as \cite{Wei2020Channel}. For the fast time-varying UT-to-RIS channel $\beamH_{2,k}$, its statistics is available, which is given by
\begin{align}
\bOmega_k = \expect { \beamH_{2,k} \odot \beamH_{2,k}^* }  \in { \R ^{ N_{\rmR} \times N_k }}.
\end{align}
The element, $\left[\bOmega_k\right]_{n,m}$, denotes the average energy coupled between the $n$th and $m$th columns of $\bU_{2,k}$ and $\bV_{2,k}$. Thus, $\bOmega_k$ is commonly referred to as the statistical eigenmode coupling matrix \cite{Gao09Statistical,You2020network}. We denote $\bx_k \in \C^{N_k \times 1} $ as the transmit signal conveyed by UT $k$, with $\bzero$ and $\bQ_k = \expect{\bx_k \bx_k^H} \in \C^{N_k \times N_k}$ being the corresponding mean vector and covariance matrix, respectively. In addition, $\bx_k$ is independent of the signals sent by other UTs, i.e., $ \expect{\bx_k \bx_{k'}^H}  = \bzero$, $\forall k' \ne k$. Then, the received signal at the BS is given by
\begin{align}
\by = \sum\nolimits_{k=1}^K { \bH_1 \bPhi \bH_{2,k} \bx_k } + \bn,
\end{align}
where $\bn \sim \mathcal{CN}(\bzero,\sigma^2 \bI_{M})$ denotes the thermal noise at the BS with variance $\sigma^2$ and the diagonal matrix $\bPhi = \diag{\phi_1,\ldots,\phi_{N_{\rmR}}}$ represents the RIS phase shift matrix.

The diagonal entry $\phi_n$, $n=1,\ldots,N_{\rmR}$, of $\bPhi$ denotes the reflection coefficient of the $n$th RIS reflecting unit. We consider an ideal RIS model with total reflection, where the reflection magnitude of each $\phi_n$ is assumed to be fixed as $\left|\phi_n\right|= 1, \forall n \in \cN$, and only the phase shifts of the reflected signals are adjustable via the RIS reflecting units. Thus, $\phi_n$ can be described by $\phi_n = e^{\jmath \theta_n},n = 1,\ldots,N_{\rmR}$, where $\theta_n$ denotes the phase shift introduced by the $n$th RIS unit. In regard to the feasible set of the phase shifts, two assumptions are considered as follows \cite{li2019sum}
\begin{itemize}
\item Continuous Phase Shift (CPS): The RIS reflecting units have infinite phase resolutions, thus can generate any desired phase values, i.e.,
    \begin{align}\label{eq:Continuous}
    \phi_n \in \cS_1 \triangleq \left\{ \phi \left| \phi = \mathrm{e}^{\jmath \theta} \right., \theta \in \left[ 0,2 \pi \right) \right\}, \forall n.
    \end{align}
\item Discrete Phase Shift (DPS): Consider a more practical RIS model with finite reflecting levels, where $\theta_n$ only can be adjusted to $\tau$ number of discrete values. For implementation convenience, we uniformly quantize the phase interval $[0,2\pi)$ to obtain available discrete phase values, i.e.,
    \begin{align}\label{eq:Discrete}
     \phi_n \in \cS_2 &\triangleq
     \left\{ \phi \left| \phi = \mathrm{e}^{\jmath \left( \frac{2\pi}{\tau}m + \frac{\pi}{\tau}\right)} \right. , m = 0, \ldots, \tau - 1 \right\}, \ntb
     &\qquad\forall n.
    \end{align}
\end{itemize}

\subsection{Average System SE and EE}
Taking expectations with respect to the UT-to-RIS channels $\bH_{2,k}$, $\forall k$, we obtain the ergodic SE expression of the RIS-aided multi-user MIMO uplink system, which is given by \cite{Wen11On}
\begin{align}\label{eq:ergodic_SE}
& \SE\left(\bQ,\bPhi\right) = \Exp \bigg\{ \log_2 \det  \bigg( \bI_M  \ntb
&\quad \ \left. \left. + \s \sum\nolimits_{k} \bH_1 \bPhi \bH_{2,k} \bQ_k \bH^H_{2,k}  \bPhi^H \bH^H_1 \right) \right\} \ [\mathrm{bits/s/Hz}],
\end{align}
where we define the collection $\bQ \triangleq \left\{ \bQ_k \right\}_{k=1}^K$. Note that $\bPhi\bH_{2,k}$ may have the same distribution as $\bH_{2,k}$ in some cases, e.g., when the entries of $\bH_{2,k}$ are independently and identically Gaussian distributed. However, in this work we focus on the jointly correlated Rayleigh fading channel model where the entries of $\bH_{2,k}$ are statistically correlated. In this case, the distribution of $\bPhi\bH_{2,k}$ would be different from that of $\bH_{2,k}$ and $\rho_{\mathrm{SE}} \left(\bQ, \bPhi\right)$ is different for different $\bPhi$ in general.

To define the EE metric, we first provide a description of the energy consumption model. Generally, the total power consumed by a RIS-empowered communication system mainly includes the transmit power at the UTs, the power consumption at the RIS, and other static hardware power required for the regular routines of the system. For further elucidation, we consider the end-to-end transmission from UT $k$ to the BS assisted by a RIS and introduce our adopted model for the corresponding $k$th single link\footnote{The single-link affine model in \eqref{eq:single} is valid under the following two conditions, namely, 1) $P_{\mathrm{c},k}$ is independent of the transmission rate; and 2) the transmit amplifiers at UTs all operate within their linear regions. For typical wireless communications, these two conditions can usually be both satisfied \cite{huang2019reconfigurable}.} as
\begin{align}\label{eq:single}
P_k = \xi_k \tr{\bQ_k} + P_{\mathrm{c},k} + P_{\mathrm{BS}} + N_{\rmR}\Ps(b),
\end{align}
where $\xi_k = \rho_k^{-1}$ with $0 < \rho_k \leq 1$ accounting for the efficiency of the transmit power amplifier adopted at UT $k$, constants $P_{\mathrm{c},k}$ and $P_{\mathrm{BS}}$ incorporate the static circuit power dissipated at UT $k$ and the BS, respectively. Lastly, the fourth term on the right hand side of \eqref{eq:single} represents the RIS energy consumption which is directly proportional to the number of RIS units. Moreover, $\Ps(b)$ denotes the per-unit hardware-dissipated power at the RIS with a $b$-bit resolution phase shifter, where the discrete variable $b$ controls the phase-shift precision as $\tau = 2^b$. In addition, we can observe from \eqref{eq:single} that the RIS operates without consuming any transmit power. As previously stated, the RIS reflectors are ideally passive, thus do not change the amplitude of the impinging signals.

Based on the above modeling of the single-link energy consumption, we have the total power consumed by the considered RIS-empowered multi-user MIMO uplink system as follows:
\begin{align}\label{eq:power_consumption}
\Psum = \sum\nolimits_{k} \left( \xi_k \tr{\bQ_k} + \Pck \right) + P_{\mathrm{BS}} + N_{\rmR} \Ps(b).
\end{align}
Then, according to the characterizations of the ergodic system SE in \eqref{eq:ergodic_SE} and the energy consumption model in \eqref{eq:power_consumption}, the average system EE is defined as
\begin{align}\label{eq:ergodic_EE}
\EE\left(\bQ,\bPhi\right) \triangleq  W\frac{\SE\left(\bQ,\bPhi\right)}{\Psum} \ [\mathrm{bits/Joule}],
\end{align}
where $W$ denotes the transmission bandwidth.

\subsection{Problem Formulation}
Instead of considering EE or SE as the only design criterion, we intend to achieve an EE-SE tradeoff via addressing a bi-criterion optimization problem \cite{Energy2013HeChunlong}, which can be efficiently tackled by the weighted sum method \cite{ehrgott2005multicriteria}, i.e., maximizing $\left(1-\alpha\right)\EE + \alpha\SE$ with $0 \leq \alpha \leq 1$. However, it seems inappropriate to add $\EE$ and $\SE$ in such a straight-forward fashion because the metric units of $\EE$ and $\SE$, which are bits/Joule and bits/s/Hz, respectively, are inconsistent. To this end, we consider a unit-consistent system metric named RE \cite{tang2014resource,you2020spectral}, which is defined as follows:
\begin{align}\label{eq:RE_definition}
\RE\left(\bQ,\bPhi\right) \triangleq \frac{\EE\left(\bQ,\bPhi\right)}{W} + \beta \frac{\SE\left(\bQ,\bPhi\right)}{\Ptot} \ [\mathrm{bits/Joule/Hz}],
\end{align}
where $\beta(>0)$ acts as the weighting factor. The denominator $\Ptot$ of the second addend is a constant and computed by
\begin{align}\label{eq:ptot}
\Ptot = \sum\nolimits_{k} \left( \Pmaxk + \Pck \right) + P_{\mathrm{BS}} + N_{\rmR} \Ps(b),
\end{align}
where $\Pmaxk$ denotes the transmit power budget at UT $k$. Accordingly, $\Ptot$ denotes the overall available power budget of the considered RIS-empowered system. Through multiplying $\EE$ and $\SE$ by the unit normalization factors $1/W$ and $1 / \Ptot$, respectively, the units of the two addends in \eqref{eq:RE_definition} are both unified into bits/Joule/Hz. Let $\beta\frac{W}{\Ptot}\triangleq \alpha/\left(1-\alpha\right)$ and substitute it into the definition of RE in \eqref{eq:RE_definition}, we can find out the equivalence between the maximization of $\RE$ and $\left(1-\alpha\right)\EE + \alpha\SE,0<\alpha<1$. Therefore, we can conclude that maximizing the RE metric is an effective way to tradeoff EE and SE with $\beta$ being the controller.

In this paper, our design objective is to exploit a reasonable EE-SE tradeoff in the RIS-aided multi-user MIMO uplink system. Based on the choice of the performance metric, we jointly design UTs' transmit covariance matrices, $\bQ_k$, $\forall k$, and the RIS phase shift matrix, $\bPhi$, to improve the system RE, which is mathematically characterized as follows:
\begin{subequations}\label{eq:problem_Q_phi}
\begin{align}
\cP_1:\quad\underset{\bQ,\bPhi} {\mathrm{maximize}} \quad & \RE\left(\bQ,\bPhi\right) \\
{\mathrm{s.t.}}\quad
& \tr { \bQ_k } \le P_{\max,k}, \quad \bQ_k \succeq \bzero,\quad\forall k \in \cK, \\
& \phi_n \in \cS ,\quad \forall n \in \cN,
\end{align}
\end{subequations}
where $\cS \in \left\{ \cS_1,\cS_2\right\}$ are defined in \eqref{eq:Continuous} and \eqref{eq:Discrete}. Notice that we can attain different EE-SE tradeoffs via properly adjusting the weight $\beta$, which is determined by system designers. Specifically, a large/small weight $\beta$ means that more emphasis should be put on the system SE/EE. In particular, the optimization of $\cP_1$ maximizes the system EE for a extremely small $\beta$, i.e., $\beta \to 0$ (with bandwidth normalization), and maximizes the system SE for a extremely large $\beta$, i.e., $\beta \to \infty$.

The optimization problem $\cP_1$ with non-convex objective and constraints is challenging to deal with. To be specific, we summarize some major difficulties in handling $\cP_1$ as follows:
\begin{description}
\item[$\mathbf{D1}$:] It is sophisticated to tackle a large number of tightly coupled variables $\bQ$ and $\bPhi$ jointly, especially for cases with a large number of RIS elements.
\item[$\mathbf{D2}$:] The calculation of the ergodic SE expression $\SE$ computing the expectation requires high-dimensional integrals, thus would incur a prohibitive computational cost.
\item[$\mathbf{D3}$:] Despite the convexity of $\SE$, the RE expression $\RE$ inherits the non-convexity from the fractional EE function $\EE$, and is even more troublesome than addressing the problem of $\EE$.
\item[$\mathbf{D4}$:] Another challenge arises from the non-convex phase-shift constraint sets, $\cS_1$ and $\cS_2$. Both cases complicate the RIS-aided RE optimization compared to those without the application of a RIS. Especially under the DPS assumption, the optimization of the RIS phase shift values is in fact a mixed integer program, which is in general non-convex.
\end{description}

In the sequel, we strive to confront the foregoing difficulties $\mathbf{D1}$-$\mathbf{D4}$ and then develop an efficient approach for handling the RE optimization problem in $\cP_1$.

\section{Optimization Framework for RE Maximization}\label{sec:optimization}

As stated in $\mathbf{D1}$, variables $\bQ$ and $\bPhi$ are nonlinearly coupled in $\cP_1$, thus being complicated to be optimized simultaneously. Therefore, we resort to the AO method to decouple the variables so that $\bQ$ and $\bPhi$ can be designed separately and sequentially. Next, we elaborate upon the main steps in the AO method, i.e., optimize $\bQ$ with $\bPhi$ fixed and optimize $\bPhi$ with $\bQ$ fixed.

\subsection{Optimization of UTs' Transmit Covariance Matrices}\label{sec:Optimization_Q}
To proceed, we first assume an arbitrarily given $\bPhi$ and then characterize the optimization of UTs' transmit covariance matrices as follows:
\begin{subequations}\label{eq:problem_Q}
\begin{align}
\cP_2:\quad\underset{\bQ} {\mathrm{maximize}} \quad & \RE\left(\bQ\right) \\
{\mathrm{s.t.}}\quad
& \tr { \bQ_k } \le P_{\max,k}, \quad \bQ_k \succeq \bzero,\quad\forall k \in \cK.
\end{align}
\end{subequations}
For further simplification, we decompose $\bQ_k$ via employing the eigenvalue decomposition, i.e.,
$\bQ_k = \bV_{k} \bLambda_k \bV_{k}^H$, where $\bV_{k} \in \C^{N_k \times N_k}$ denotes the eigenmatrix with its columns being the eigenvectors of $\bQ_k$, and $\bLambda_k \in \R^{N_k \times N_k}$ denotes the power allocation matrix with the eigenvalues of $\bQ_k$ along its main diagonal. In fact, $\bV_k$ represents the transmit subspace at UT $k$ consisting of signal directions and $\bLambda_k$ denotes the corresponding power allocated to each direction. From the eigenvalue decomposition, we can find that handling $\cP_2$ is equivalent to optimizing $\bV_k$, $\forall k$, and $\bLambda_k$, $\forall k$. In the following, we will respectively seek for $\bV_k$, $\forall k$, and $\bLambda_k$, $\forall k$.

\subsubsection{Optimal Transmit Directions at UTs}
Firstly, the optimal transmit subspace at the UT sides can be analytically known as detailed in the following proposition.
\begin{prop}\label{prop:beam_domain_optimal}
The optimal eigenmatrix $\bV_{k}$ of $\bQ_k$ is identical to the unitary matrix $\bV_{2,k}$, which is defined in the decomposition of $\bH_{2,k}$ in \eqref{eq:beam_H2}, i.e.,
\begin{align}
\bV_{k} = \bV_{2,k}, \quad \forall k.
\end{align}
\end{prop}
The proposition can be proved by following an approach as in e.g., \cite{you2020energy}, thus being omitted here for brevity.

\propref{prop:beam_domain_optimal} indicates that to obtain the maximum system RE in the considered RIS-aided MIMO uplink system, the optimal transmit directions at UT $k$ should be within the signal space spanned by the eigenvectors of the corresponding channel's transmit correlation matrix. Thereby, we can determine the signal directions following the result in \propref{prop:beam_domain_optimal}, and then focus on the design of power allocation strategies at the UT sides. Mathematically, by setting $\bV_{k} = \bV_{2,k}$, $\forall k$, the precoding design problem in $\cP_2$ can be transformed to a power allocation problem as follows:
\begin{subequations}\label{eq:problem_Lambda}
\begin{align}
\cP_3: \quad\underset{\bLambda} {\mathrm{maximize}} \quad & f_3 \left(\bLambda\right) = \frac{\EEt\left(\bLambda\right)}{W} + \beta \frac{\SEt\left(\bLambda\right)}{\Ptot}\\
{\mathrm{s.t.}}\quad
& \tr { \bLambda_k } \le \Pmaxk, \; \bLambda_k \succeq \bzero, \ntb
& \bLambda_k \; \mathrm{diagonal},\; \forall k\in \cK,
\end{align}
\end{subequations}
where $\bLambda \triangleq \left\{ \bLambda_k \right\}_{k=1}^K$ and
\begin{align}  \label{eq:EEDE}
& \EEt\left(\bLambda\right) = W\frac{\SEt\left(\bLambda\right)}{P\left(\bLambda\right)},\\ \label{eq:SEDE}
& \SEt\left(\bLambda\right) = \Exp \bigg\{ \log_2 \det \bigg( \bI_M \ntb
& \quad \left. \left. + \s \sum\nolimits_{k} \bH_1 \bPhi \bU_{2,k} \beamH_{2,k} \bLambda_k \beamH^H_{2,k} \bU^H_{2,k} \bPhi^H \bH^H_1 \right) \right\},\\
& P\left(\bLambda\right) = \sum\nolimits_{k} \left( \tr{\bLambda_k} + \Pck \right) + P_{\mathrm{BS}} + N_{\rmR} \Ps(b).
\end{align}

\subsubsection{Asymptotic System SE}
Compared with $\cP_2$, the number of variables in $\cP_3$ has been dramatically reduced. Nonetheless, $\cP_3$ is still intractable due to the burdensome computation of the expectation values, as is stated in $\mathbf{D2}$. In fact, stochastic programming can be adopted to address $\cP_3$. However, without an analytical expression of the objective function, the execution time would be exceedingly long since Monte-Carlo methods rely on exhaustive channel averaging in each iteration. To lift the computational burden, we derive an asymptotic approximation of the objective called deterministic equivalent (DE) via leveraging the random matrix theory \cite{Couillet11Random}. To this end, we consider a large-scale MIMO system where $M$ and $N_k$, $\forall k$, both tend to infinity with the ratios $c_k = M / N_k$, $\forall k$, fixed. Rewriting \eqref{eq:SEDE} in a compact form as follows:
\begin{align}\label{eq:R_bigLambda}
\SEt\left(\bLambda\right) & = \expect{\log_2\det\left(\bI_M + \s \bG \bD \bG^H \right)},
\end{align}
where $\bD = \diag{\bLambda_1,\cdots,\bLambda_K} \in \R^{N \times N}$, $\bG = \left[ \bG_1  \cdots \bG_K  \right] \in \C^{ M \times N}$, $\bG_k = \bH_1 \bPhi \bU_{2,k} \beamH_{2,k} \in \C^{ M \times N_k}$, $\forall k$, and $N = \sum\nolimits_k{N_k}$. Then, following the results in \cite{Wen11On,you2020reconfigurable}, an asymptotic expression of $\SEt\left(\bLambda\right)$ is given by
\begin{align}\label{eq:DE}
\SEde\left(\bLambda\right) & = \sum\nolimits_{k} \log_2 \det \left( \bI_{N_k} + \bGamma_k \bLambda_k \right) \ntb
& \quad + \log_2 \det \left( \bI_M + \bPsi \right) - \sum\nolimits_{k} {\bgamma_k^T \bOmega_k \bpsi_k},
\end{align}
where $\bgamma_k \triangleq \left[ \gamma_{k,1},\ldots,\gamma_{k,N_{\rmR}} \right]^T$, $\bpsi_k \triangleq \left[\psi_{k,1},\ldots,\psi_{k,N_k} \right]^T$, and $\bPsi \triangleq \sum\nolimits_k{\bPsi_k} \in \C^{M \times M}$.
Define $\bU_{\rmG_k} \triangleq \bH_1 \bPhi \bU_{2,k} \in \C^{ M \times N_{\rmR}}$, $\forall k$, we calculate $\bGamma_k$ and $\bPsi_k$ by
\begin{align}\label{eq:T}
\bGamma_k & =  \diag{ \bOmega_k^T \bgamma_k } \in \C^{N_k \times N_k}, \\ \label{eq:F}
\bPsi_k & = \s \bU_{\rmG_k} \diag{ \bOmega_k \bpsi_k } \bU_{\rmG_k}^H \in \C^{ M \times M },
\end{align}
respectively. Lastly, the DE auxiliary quantities $\bgamma \triangleq \left\{\gamma_{k,m}\right\}_{\forall k,m}$ and $\bpsi \triangleq \left\{\psi_{k,n}\right\}_{\forall k,n}$ are the unique solutions to the following iterative equations:
\begin{align}\label{eq:gamma}
\gamma_{k,m} & = \s \bu_{\rmG_k,m}^H \left( \bI_M + \bPsi \right)^{-1} \bu_{\rmG_k,m},\,  \forall m \in \cN, \, \forall k \in \cK, \\ \label{eq:psi}
\psi_{k,n} & = \frac{\lambda_{k,n,n}}{1+g_{k,n,n} \lambda_{k,n,n}}, \, \forall n \in \cN_k, \, \forall k \in \cK,
\end{align}
where $\cN_k = \left\{1,\ldots,N_k\right\}$ and $\bu_{\rmG_k,m}$ denotes the $m$th column of $\bU_{\rmG_k}$. Moreover, $\lambda_{k,n,n}$ and $g_{k,n,n}$ are the $(n,n)$th elements of $\bLambda_k$ and $\bGamma_k$, respectively. Given an initial point of $\bgamma_k$ or $\bpsi_k$, we can easily obtain the fixed-point solutions $\bgamma$ and $\bpsi$ via cyclically updating them by \eqref{eq:gamma} and \eqref{eq:psi}.

Notice that although derived under the assumption of large-scale MIMO systems, the asymptotic approximation in \eqref{eq:DE} is still sufficiently accurate for characterizing the ergodic SE of small-scale MIMO systems \cite{you2020reconfigurable,Wen11On}. To summarize, the proposed DE method obtaining \eqref{eq:DE} is given in \alref{alg:Deterministic}.

\begin{algorithm}[h]
\caption{DE Method.}
\label{alg:Deterministic}
\begin{algorithmic}[1]
\State Initialize threshold $\varepsilon$.
\For{$k=1$ to $K$}
\State Initialize $\bpsi_k^{(0)}$ and set iteration index $u = 0$.
\Repeat
\For{$m=1$ to $M$}
\State Update $\gamma_{k,m}^{(u+1)}$ by \eqref{eq:gamma} with $\bpsi_k^{(u)}$.
\EndFor
\For{$n=1$ to $N_k$}
\State Update $\psi_{k,n}^{(u+1)}$ by \eqref{eq:psi} with $\bgamma^{(u+1)}_k$.
\EndFor
\State Set $u=u+1$.
\Until{$\left|  {\bpsi}_k^{(u)} -  {\bpsi}_k^{(u-1)} \right|\le \varepsilon$.}
\State Calculate $\bGamma_k$ and $\bPsi_k$ with the aid of ${\bgamma}_k^{(u)}$ and ${\bpsi}_k^{(u)}$ by \eqref{eq:T} and \eqref{eq:F}, respectively.
\EndFor
\State Calculate the DE expression $\SEde\left(\bLambda\right)$ in \eqref{eq:DE} with the aid of $\bGamma_k$ and $\bPsi_k$, $\forall k$.
\Ensure The DE auxiliary quantities $\bpsi = \bpsi^{(u)}$, and DE-based system SE, $\SEde\left(\bLambda\right)$.
\end{algorithmic}
\end{algorithm}

Then, by means of replacing $\SEt\left(\bLambda\right)$ in the objective function of $\cP_3$ with its asymptotic expression $\SEde \left( \bLambda \right)$, we arrive at an asymptotic optimization program as follows:
\begin{subequations}\label{eq:problem_Lambda_DE}
\begin{align}\label{eq:objective}
\overline{\cP}_3: \quad\underset{\bLambda} {\mathrm{maximize}} \quad & \overline{f}_3 \left(\bLambda\right)
= \frac{\SEde\left(\bLambda\right)}{P\left(\bLambda\right)} + \beta \frac{\SEde\left(\bLambda\right)}{\Ptot} \\
{\mathrm{s.t.}}\quad
& \tr { \bLambda_k } \le \Pmaxk, \; \bLambda_k \succeq \bzero,\ntb
& \bLambda_k \; \mathrm{diagonal},\; \forall k\in \cK.
\end{align}
\end{subequations}

Instead of requiring the knowledge of the actual realizations of $\beamH_{2,k}$, $\forall k$, to compute the ergodic SE in \eqref{eq:ergodic_SE}, the asymptotic SE in \eqref{eq:DE} can be computed with only the statistics of $\beamH_{2,k}$, $\forall k$. Therefore, the DE-based optimization problem $\overline{\cP}_3$ enables us to design transmit strategies that exploit the statistical knowledge of the UT-to-RIS channels. Moreover, since the objective of $\overline{\cP}_3$ is actually a function of $(\bgamma,\bpsi)$, we update $\bLambda$ and $(\bgamma,\bpsi)$ in an iterative manner, as adopted in \cite{Wen11On}. In addition, the solution to $\overline{\cP}_3$ is almost surely accurate since $\SEde\left(\bLambda\right)$ is asymptotically accurate as the matrix sizes of $\bH_{2,k}$, $\forall k$, tend to infinity \cite{Wen11On,Couillet11Random}.

\subsubsection{Quadratic Transformation}
It is worth noting that the asymptotic system SE, $\SEde\left(\bLambda\right)$, is a strictly concave function with respect to $\bLambda$ \cite{Wen11On}. Nevertheless, $\overline{\cP}_3$ is still generally non-concave due to the existence of a fractional term in its objective function. Fortunately, the fractional term, $\frac{\SEde\left(\bLambda\right)}{P\left(\bLambda\right)}$, is concave-convex so that the numerator and the denominator can be decoupled via applying the quadratic transformation \cite{shen2018fractional}. Specifically, by introducing a auxiliary variable $y\in\R$, we can equivalently convert $\overline{\cP}_3$ into a non-fractional problem as follows:
\begin{subequations}\label{eq:Quadratic}
\begin{align}
\cP_4: \quad\underset{\bLambda,y} {\mathrm{maximize}} \quad & f_4 \left(\bLambda,y\right) = 2y\sqrt{\SEde\left(\bLambda\right)} - y^2P\left(\bLambda\right) \ntb
& \quad\quad\quad\quad\quad + \beta \frac{\SEde\left(\bLambda\right)}{\Ptot}\\
{\mathrm{s.t.}}\quad
& \tr { \bLambda_k } \le \Pmaxk, \; \bLambda_k \succeq \bzero,\ntb
& \bLambda_k \; \mathrm{diagonal},\; \forall k\in \cK.
\end{align}
\end{subequations}
To address $\cP_4$, we optimize the variables $\bLambda$ and $y$ in a separate and iterative manner. With an arbitrarily given $\bLambda$, the optimal $y$ can be directly obtained as
\begin{align}\label{eq:optimal_y}
y^* = \frac{\sqrt{\SEde\left(\bLambda\right)}}{P\left(\bLambda\right)}.
\end{align}

Then, we consider the optimization of the primal variable $\bLambda$ with a given $y$. It is worth recalling that $\SEde \left( \bLambda\right)$ is a concave function so that the square-root function $\sqrt{ \SEde \left( \bLambda \right)}$ is concave and nondecreasing. Furthermore, the term $-P\left( \bLambda \right)$ is also concave. Based on these facts, we can find that $\cP_4$ is concave over $\bLambda$ for a fixed $y$. Hence, $\cP_4$ can be numerically and efficiently solved via using classical convex optimization \cite{Boyd04Convex}. More explicit descriptions of the iterative method for solving $\overline{\cP}_3$ based on the quadratic transformation are summarized in \alref{alg:Quadratic}.

\begin{algorithm}[h]
\caption{Quadratic Transformation for Solving $\overline{\cP}_3$.}
\label{alg:Quadratic}
\begin{algorithmic}[1]
\Require Feasible $\bLambda^{(0)}$, iterative index $q=0$, and threshold $\varepsilon$.
\Repeat
\State Calculate $y^{(q+1)}$ via using $\bLambda^{(q)}$ by \eqref{eq:optimal_y}.
\State Update $\bLambda^{(q+1)}$ via solving the reformulated problem $\cP_4$ with fixed $y=y^{(q+1)}$.
\State Update the objective value $\overline{f}_3 \left(\bLambda^{(q+1)}\right)$ via using $\bLambda^{(q+1)}$.
\State Set $q = q + 1$.
\Until{$\left| \overline{f}_3 \left(\bLambda^{(q)}\right) - \overline{f}_3 \left(\bLambda^{(q-1)}\right) \right| \le \varepsilon$.}
\Ensure The solution $\bLambda^* = \bLambda^{(q+1)}$ to $\overline{\cP}_3$.
\end{algorithmic}
\end{algorithm}

\begin{prop}
According to the convergence properties stated in \cite[Theorem 3]{shen2018fractional}, \alref{alg:Quadratic} can generate a convergent and non-decreasing sequence of the objective values of $\overline{\cP}_3$. In addition, the final result $\bLambda^*$ output by \alref{alg:Quadratic} is a stationary point of $\overline{\cP}_3$.
\end{prop}

\subsection{Adjustment of RIS Phase Shifters}\label{sec:Optimization_Phi}
In this subsection, we discuss the other indispensable step in the AO method, i.e., designing the phase shifts for the RIS elements with UTs' transmit covariance matrices being fixed. Notice that the value of the total energy consumption is independent of $\bPhi$ because no transmit power is consumed by the RIS. Therefore, assuming $\bQ$ is fixed, optimizing $\bPhi$ for RE maximization is equivalent to that for the maximization of SE. Inspired by this fact, we simplify the optimization of $\bPhi$ to that maximizes the system SE only. In addition, the optimizations of $\bPhi$ and the DE auxiliary parameters $(\bgamma,\bpsi)$ are executed successively and cyclically \cite{Wen11On}, i.e., we fix $(\bgamma,\bpsi)$ when updating $\bPhi$ and then adjust $(\bgamma,\bpsi)$ by \eqref{eq:gamma} and \eqref{eq:psi}. Let us consider the DE expression in \eqref{eq:DE} and treat $\bgamma$ and $\bpsi$ as constants. Then, only the second term, $\log_2 \det \left( \bI_M + \bPsi \right)$, in \eqref{eq:DE} is associated with $\bPhi$, while the others are irrelevant to $\bPhi$. Dropping all the constant terms, the optimization of $\bPhi$ can be boiled down to the following problem:
\begin{subequations}\label{eq:problem_Phi}
\begin{align}
\cP_5: \quad \underset{\bPhi} {\mathrm{maximize}} \quad & f_5 \left( \bPhi \right) = \log_2 \det \left( \bI_M \right. \ntb
 + \sum\nolimits_{k} \s & \left. \bH_1 \bPhi \bU_{2,k} \diag{ \bOmega_k \bpsi_k } \bU_{2,k}^H \bPhi^H \bH_1^H \right)\\
{\mathrm{s.t.}}\quad
& \phi_n \in \cS ,\quad \forall n \in \cN.
\end{align}
\end{subequations}
Despite the simplification of the objective function, it is still arduous to straightforwardly handle $\cP_5$ for two reasons. For one thing, the objective is non-convex with respect to $\bPhi$ and thus is intractable. Besides, the constraints on the RIS phase shifters are intractable. To be specific, for the CPS case, $\cS_1$ is non-convex and is a manifold where every $\bphi_n$ is restricted to be unit-modulus. For the case of DPS, $\cS_2$ is discrete and also non-convex, such that $\cP_5$ is in general a mixed integer program. In the sequence, we will focus on tackling these technical challenges and then build an efficient approach to address $\cP_5$.

\subsubsection{Weighted Minimum MSE (WMMSE) Method}
To circumvent the intractable function in the objective of $\cP_5$, we resort to the WMMSE method, which equivalently converts a challenging SE maximization problem into an MSE minimization problem with a more tractable form \cite{shi2011an}. To proceed, we define $\bA=\sum\nolimits_{k}{ \bU_{2,k} \diag{ \bOmega_k \bpsi_k } \bU_{2,k}^H } \succeq \bzero$ for notational conciseness. Regard $f_5 \left( \bPhi \right)$ in \eqref{eq:problem_Phi} as the system SE of a hypothetical communication system with $\bH_1$ being the channel matrix, $\bs_{\mathrm{h}} \sim \mathcal{CN}(\bzero,\bI_{N_{\rmR}})$ being the signal symbol, $\bPhi \bA^{1/2}$ being the hypothetical beamformer, and $\bU_{\mathrm{h}} \in \C^{M \times N_{\mathrm{R}}}$ being the linear receiving matrix. Then, the MSE matrix is given as follows:
\begin{align}\label{eq:MSE}
\bE_{\mathrm{h}} & = \left( \bU_{\mathrm{h}}^H \bH_1 \bPhi \bA^{1/2} - \bI_{N_{\mathrm{R}}} \right)\left( \bU_{\mathrm{h}}^H \bH_1 \bPhi \bA^{1/2} - \bI_{N_{\mathrm{R}}} \right)^H \ntb
& \quad\quad + \sigma^2 \bU_{\mathrm{h}}^H \bU_{\mathrm{h}}.
\end{align}
Based on these hypotheses and then applying the WMMSE method, $\cP_5$ can be transformed into the following MSE minimization problem:
\begin{subequations}\label{eq:MMSE}
\begin{align}
\cP_{5a}:\quad \underset{\bW_{\mathrm{h}},\bU_{\mathrm{h}},\bPhi} {\mathrm{minimize}} \quad & f_{5a} \left( \bW_{\mathrm{h}},\bU_{\mathrm{h}},\bPhi \right) \triangleq \tr{\bW_{\mathrm{h}} \bE_{\mathrm{h}}} \ntb
& \quad - \log_2 \det \left( \bW_{\mathrm{h}} \right) \\
{\mathrm{s.t.}}\quad
& \phi_n \in \cS ,\quad \forall n \in \cN,
\end{align}
\end{subequations}
where $\bW_{\mathrm{h}}\in \C^{N_{\mathrm{R}} \times N_{\mathrm{R}}}$ is an auxiliary variable. The equivalence between the solutions of $\bPhi$ to $\cP_5$ and $\cP_{5a}$ is guaranteed \cite[Theorem 1]{shi2011an}. In addition, the latter problem $\cP_{5a}$ is more convenient to tackle since the objective function is convex over each variable ($\bW_{\mathrm{h}}$, $\bU_{\mathrm{h}}$ or $\bPhi$) while holding others fixed. Next, exploiting the aforementioned equivalence relationship as well as the convexity property, we will develop a computationally-efficient iterative WMMSE algorithm for the SE maximization problem $\cP_5$.

\subsubsection{Block Coordinate Descent (BCD) Method}
To decouple the large number of variables in $\cP_{5a}$, we handle it via applying the BCD method, which is one of the fundamental approaches for tackling large-size optimization problems. The optimization is accomplished by an iterative procedure where we separately update one variable while treating the other two as constants. The solutions of $\bW_{\mathrm{h}}$ and $\bU_{\mathrm{h}}$ are clear and explicit, which are respectively given by
\begin{align}\label{eq:optimal_W}
\bW_{\mathrm{h}}^{\mathrm{opt}} & = \bE_{\mathrm{h}}^{-1}, \\ \label{eq:optimal_U}
\bU_{\mathrm{h}}^{\mathrm{opt}} & = \left( \sigma^2 \bI_M + \bH_1 \bPhi \bA \bPhi^H \bH_1^H \right)^{-1}\bH_1 \bPhi \bA^{1/2}.
\end{align}
The crux of our approach lies in the adjustment of $\bPhi$, which is formulated as follows:
\begin{subequations}\label{eq:MSE_Phi}
\begin{align}
\cP_6:\quad \underset{\bPhi} {\mathrm{minimize}} \quad & f_6 \left(\bPhi\right) = \tr{ \bPhi^H \bB \bPhi \bA  } \ntb
& \quad - \tr{\bPhi^H \bC^H } - \tr{\bPhi \bC } \\
{\mathrm{s.t.}}\quad
& \phi_n \in \cS ,\quad \forall n \in \cN,
\end{align}
\end{subequations}
where we remove the constant terms irrelevant to $\bPhi$ and define $\bB = \bH_1^H \bU_{\mathrm{c}} \bW_{\mathrm{c}} \bU_{\mathrm{c}}^H \bH_1 \in \C^{N_{\mathrm{R}} \times N_{\mathrm{R}}}$ and $\bC = \bA^{1/2}\bW_{\mathrm{c}}\bU_{\mathrm{c}}^H \bH_1 \in \C^{N_{\mathrm{R}} \times N_{\mathrm{R}}}$ for notational conciseness. Denoting $\bphi \triangleq \left[\phi_1,\ldots,\phi_{N_{\rmR}}\right]^T$, we have $\tr{ \bPhi^H \bB \bPhi \bA  } = \bphi^H \left( \bB \odot \bA^T \right) \bphi$, which follows from the matrix identity in \cite[Eq. (1.10.6)]{Zhang2017Matrix}. Then, $\cP_6$ can be further simplified into
\begin{subequations}\label{eq:MSE_phi}
\begin{align}
\cP_{6a}:\quad \underset{\bphi} {\mathrm{minimize}} \quad & f_{6a} \left(\bphi\right) = \bphi^H \left( \bB \odot \bA^T \right) \bphi - 2 \Real{\bphi^H\bc^*}  \\
{\mathrm{s.t.}}\quad
& \phi_n \in \cS ,\quad \forall n \in \cN,
\end{align}
\end{subequations}
where $\bc \triangleq \left[ \left[\bC\right]_{1,1}, \ldots, \left[\bC\right]_{N_{\mathrm{R}},N_{\mathrm{R}}} \right]^T$.

The main procedure of the iterative WMMSE method for addressing $\cP_5$ is presented in \alref{alg:BCD}. However, there remains a problem in step 6 of \alref{alg:BCD} that has to be addressed. In the sequel, we concentrate on tackling this issue for updating $\bphi$ in each BCD iteration.

\begin{algorithm}[h]
\caption{Iterative WMMSE Method for $\cP_5$.}
\label{alg:BCD}
\begin{algorithmic}[1]
% \State Initialize:
\Require Feasible $\bW_{\mathrm{h}}^{(0)}$, $\bU_{\mathrm{h}}^{(0)}$, $\bPhi^{(0)}$, iterative index $s=0$.
\Repeat
\State Update $\bU_{\mathrm{h}}^{(s+1)}$ with $\bPhi^{(s)}$ by \eqref{eq:optimal_U}.
\State Calculate $\bE_{\mathrm{h}}^{(s+1)}$ with $\bU_{\mathrm{h}}^{(s+1)}$ and $\bPhi^{(s)}$ by \eqref{eq:MSE}.
\State Update $\bW_{\mathrm{h}}^{(s+1)} = \left(\bE_{\mathrm{h}}^{(s+1)}\right)^{-1}$.
\State Update $\bphi^{(s+1)}$ with $\bW_{\mathrm{h}}^{(s+1)}$ and $\bU_{\mathrm{h}}^{(s+1)}$ by solving $\cP_{6a}$.
\State Update $\bPhi^{(s+1)} = \diag{\bphi^{(s+1)}}$.
\State Set $s = s + 1$.
%\Until{$\left|  f_{5a} \left( \bW_{\mathrm{h}}^{(s)},\bU^{(s)}_{\mathrm{h}},\bPhi^{(s)} \right) -  f_{5a} \left( \bW_{\mathrm{h}}^{(s-1)},\bU_{\mathrm{h}}^{(s-1)},\bPhi^{(s-1)} \right) \right|\le \varepsilon$}
\Until{Some stopping criterion is satisfied.}
\Ensure The solution $\bPhi^* = \bPhi^{(s)}$ to $\cP_5$.
\end{algorithmic}
\end{algorithm}

\subsubsection{Penalty Method}
Note that $\cP_{6a}$ is essentially non-convex even though its objective function is convex, where the non-convexity arises from constraints on the phase shifts. Our strategy for dealing with these intractable constraints relies on a penalty method called negative square penalty (NSP) method, which is in essence an application of the homotopy optimization approach \cite{shao2019a,shao2020binary}. The idea is to approximate a challenging problem by an easy-to-handle one via imposing a penalty on the objective function. To proceed, we first introduce the following proposition.
\begin{prop}\label{prop:prop3}
Consider a minimization problem with a general form as follows:
\begin{align}\label{eq:NSP1}
\cF1: \quad \underset{\bphi} {\mathrm{minimize}} \quad & f(\bphi) \ntb
{\mathrm{s.t.}}\quad
& \phi_n \in \cS ,\quad \forall n \in \cN.
\end{align}
Imposing a negative square penalty on $f(\bphi)$ with $\lambda > 0$ being the penalty parameter, we consider the following penalty reformulation of $\cF1$ as
\begin{align}\label{eq:NSP2}
\cF2: \quad \underset{\bphi} {\mathrm{minimize}} \quad & f(\bphi) - \lambda \left\|\bphi\right\|^2 \ntb
{\mathrm{s.t.}}\quad
& \phi_n \in \widetilde{\cS},\quad \forall n \in \cN,
\end{align}
where $\widetilde{\cS}$ represents the convex hull of $\cS$ and we define $\widetilde{\cS}^{N_{\rmR}} \triangleq \left\{ \bphi \in \C^{N_{\rmR}} \left| \phi_n \in \widetilde{\cS} \right.,\forall n \right\}$. Assume that $f:\C^{N_{\rmR}} \to \R$ is Lipschitz continuous on $\widetilde{\cS}^{N_{\rmR}}$, i.e., $\left| f(\bx)-f(\by)\right| \leq L \left\| \bx - \by \right\|, \forall \, \bx,\by \in \widetilde{\cS}^{N_{\rmR}}$, where $L > 0$ represents the Lipschitz constant. Then, a constant $\overline{\lambda} > 0$ exists such that for any $\lambda > \overline{\lambda}$, if $\bx^*$ is (globally) optimal for $\cF1$, it is also (globally) optimal for $\cF2$. The inverse also holds. The threshold hold values of $\overline{\lambda}$ are $\overline{\lambda} = L$ and $\overline{\lambda} = L / \sin(\pi / \tau)$ for the
cases with CPS and DPS, respectively \cite[Theorem 1]{shao2019a}.
\end{prop}

It is worth noting that the exact penalty method widely adopted in nonlinear programming \cite{nocedal2006numerical} is also applicable to the minimization problem $\cF1$ with non-convex constraints. For instance, the exact penalty method could transform $\cF1$ for the case of DPS into an unconstrained problem as follows:
\begin{align}\label{eq:PM}
\underset{\bphi} {\mathrm{minimize}} \quad  f(\bphi) + \lambda \sum\nolimits_{n=1}^{N_{\rmR}}{\left| 1 - \left(\phi_n \mathrm{e}^{-\jmath \frac{\pi}{\tau}} \right)^\tau \right|}.
\end{align}
The penalty function in \eqref{eq:PM} forces $\left(\phi_n \mathrm{e}^{-\jmath \frac{\pi}{\tau}} \right)^\tau = 1$, such that $\phi_n \in \cS_2$, $\forall n$. However, the higher-order polynomials involved in the objective function are also challenging to deal with. In comparison, the penalty function in the NSP method is quadratic and is independent of the dimension of $\bphi$, i.e., $N_{\rmR}$. Therefore, we choose the NSP method to handle $\cP_{6a}$. Specifically, accordingly to \propref{prop:prop3}, by choosing a proper penalty parameter $\lambda$ satisfying the equivalence condition, we reformulate $\cP_{6a}$ into the following problem:
\begin{subequations}\label{eq:NSP}
\begin{align}
\cP_{7}:\quad \underset{\bphi} {\mathrm{minimize}} \quad & f_{\lambda} \left(\bphi\right) = f_{6a}\left(\bphi\right) - \lambda \left\|\bphi\right\|^2 \\
{\mathrm{s.t.}}\quad
& \phi_n \in \widetilde{\cS},\quad \forall n \in \cN.
\end{align}
\end{subequations}
It is worth mentioning that to ensure the equivalence between $\cP_{6a}$ and $\cP_{7}$, finely adjusting $\lambda$ is unnecessary, as an adequately large $\lambda$ is sufficient to obtain the equivalence.

Recall that $\widetilde{\cS}$ is the convex hull of $\cS$. Therefore, for the case of $\cS_1$, $\widetilde{\cS}_1$ is a unit circle, i.e., $\widetilde{\cS}_1 = \left\{ \phi \in \C \left| \left|\phi\right| \leq 1 \right. \right\}$, and for the case of $\cS_2$, $\widetilde{\cS}_2$ is a regular polygon with vertices $\left\{1, \mathrm{e}^{\jmath \frac{2\pi}{\tau}}, \ldots, \mathrm{e}^{\jmath \frac{2\pi\left(\tau-1\right)}{\tau}}\right\}$ \cite{li2019sum}. Since the feasible set $\widetilde{\cS}$ of $\cP_{7}$ is convex and exhibits a nice geometric structure for both $\cS_1$ and $\cS_2$, $\cP_{7}$ is much more manageable than the primal non-convex-constrained problem $\cP_{6a}$. In the following, we will propose an efficient algorithm for addressing $\cP_{7}$.

\subsubsection{Gradient Extrapolated MM Method}
The functions $f_{6a}(\bphi)$ and $\lambda \left\|\bphi\right\|^2$ in \eqref{eq:NSP} are both convex and quadratic. Thus, problem $\cP_{7}$ is a non-convex program with its objective function $f_{\lambda} \left(\bphi\right)$ exhibiting a typical difference-of-convex-functions form. We can handle $\cP_{7}$ by applying the MM method, which is a classical sequential optimization technique. Generally, the MM method hinges on constructing a majorizer of $f_{\lambda} (\bphi)$. More specifically, we have to find a majorant function $F_{\lambda}(\bphi \mid \overline{\bphi})$ to approximate $f_{\lambda} (\bphi)$ at point $\overline{\bphi}$ and $F_{\lambda}(\bphi \mid \overline{\bphi})$ should satisfy the following three conditions:
\begin{description}
\item[$\mathbf{C1}$:] $F_{\lambda}(\bphi \mid \overline{\bphi}) \geq f_{\lambda} (\bphi),\ \forall \bphi, \overline{\bphi} \in \widetilde{\cS}$,
\item[$\mathbf{C2}$:] $F_{\lambda}(\overline{\bphi}\mid \overline{\bphi}) = f_{\lambda} (\overline{\bphi}),\ \forall \overline{\bphi} \in \widetilde{\cS}$,
\item[$\mathbf{C3}$:] $\nabla_{\bphi} \ F_{\lambda}(\bphi \mid \overline{\bphi}) = \nabla_{\bphi} \ f_{\lambda} (\bphi),\ \forall \bphi,\overline{\bphi} \in \widetilde{\cS}$.
\end{description}
To derive a majorant function for the problem at hand, we consider the inequality $\left\|\bphi\right\|^2 \geq \left\|\overline{\bphi}\right\|^2 + 2 \langle \overline{\bphi}, \bphi - \overline{\bphi}\rangle$. Accordingly, we have
\begin{align}\label{eq:surrogate}
f_{\lambda} (\bphi) \leq F_{\lambda}(\bphi \mid \overline{\bphi}) = f_{6a}\left(\bphi\right) - \lambda \left( \left\|\overline{\bphi}\right\|^2 + 2 \langle \overline{\bphi}, \bphi - \overline{\bphi}\rangle \right).
\end{align}
It is not difficult to check the qualification of $F_{\lambda}(\bphi \mid \overline{\bphi})$ as a majorant function of $f_{\lambda} (\bphi)$, since $F_{\lambda}(\bphi \mid \overline{\bphi})$ satisfies all the conditions, $\mathbf{C1}$-$\mathbf{C3}$. Then, the MM method for addressing $\cP_{7}$ iteratively performs the optimization as follows:
\begin{subequations}\label{eq:MM}
\begin{align}
\cP_{8}:\quad \bphi^{(\ell+1)} = \arg \underset{\bphi} \min \quad & F_{\lambda}(\bphi \mid \bphi^{(\ell)})  \\
{\mathrm{s.t.}}\quad
& \phi_n \in \widetilde{\cS},\quad \forall n \in \cN.
\end{align}
\end{subequations}
Note that the majorant function $F_{\lambda}(\bphi \mid \bphi^{(\ell)})$ linearizes the concave term $- \lambda \left\|\bphi\right\|^2$ in $f_{\lambda} (\bphi)$ at the point $\bphi^{(\ell)}$, which is the minimizer in the $\ell$th iteration. Consequently, $\cP_{8}$ is convex and also smooth under the constructed convex set $\widetilde{\cS}$. The projected gradient (PG) method, as well as the accelerated projected gradient (APG) method, can be employed to solve smooth convex problems and is especially suitable when it is easy to calculate the projection operators \cite{beck2017first}. Moreover, the APG method has a faster convergence rate compared to the PG method for cases of convex problems \cite{shao2019a}. Therefore, we choose the APG method herein to find a solution to $\cP_{8}$. Specifically, utilizing the negative gradient, the APG method solves $\cP_{8}$ by iteratively updating
\begin{align}\label{eq:APG}
\bx^{(i+1)} = \Pi_{\widetilde{\cS}^{N_{\rmR}}}\left(\bz^{(i)} - \frac{1}{\beta_{(i)}}\nabla_{\bphi}F_{\lambda}(\bz^{(i)} \mid \bphi^{(\ell)})\right),
\end{align}
where $i$ is the iterative index, $\Pi_{\widetilde{\cS}^{N_{\rmR}}}(\bphi) = \arg \underset{\hat\bphi\in\widetilde{\cS}^{N_{\rmR}}} \min \left\|\bphi-\hat\bphi\right\|^2$ denotes the projection of $\bphi$ onto $\widetilde{\cS}^{N_{\rmR}}$, $\frac{1}{\beta_{(i)}}$ and $\bz^{(i)}$ denote the step length and the extrapolated point at the $i$th iteration, respectively. We compute $\bz^{(i)}$ by
\begin{align}\label{eq:extrapolated}
\bz^{(i)} = \bx^{(i)} + \alpha_{(i)}(\bx^{(i)} - \bx^{(i-1)}),
\end{align}
where
\begin{subequations}\label{eq:alpha}
\begin{align}
\alpha_{(i)} & = \frac{\zeta_{(i-1)}-1}{\zeta_{(i)}},\\
\zeta_{(i)} & = \frac{1+\sqrt{1+4\zeta_{(i-1)}^2}}{2},
\end{align}
\end{subequations}
with initialization $\bx^{(-1)} = \bx^{(0)}$ and $\zeta_{(-1)} = 0$. In addition, the step length $\frac{1}{\beta_{(i)}}$ should be properly chosen such that $\bx^{(i+1)}$ meets the following property:
\begin{align}\label{eq:descent property}
& F_{\lambda}(\bx^{(i+1)} \mid \hat\bx)  \leq F_{\lambda}(\bz^{(i)} \mid \hat\bx) + \frac{\beta_{(i)}}{2}\left\|\bx^{(i+1)}-\bz^{(i)}\right\|^2 \ntb
& \quad + \left\langle \nabla_{\bphi}F_{\lambda}(\bz^{(i)} \mid \hat\bx), \bx^{(i+1)}-\bz^{(i)}\right\rangle, \; \forall \hat\bx \in \widetilde{\cS}^{N_{\rmR}}.
\end{align}

It is worth noting that to exactly solve every MM subproblem in $\cP_{8}$, multiple APG iterations are required, which could create high computational burden. In order to reduce the computational complexity, our strategy is to run only one round of APG iteration in each MM update, which is referred to as the gradient extrapolated MM (GEMM) method \cite{li2016sum}. More specifically, the inexact GEMM method with one-step APG update is given by
\begin{align}\label{eq:GEMM}
\bphi^{(\ell+1)} & = \Pi_{\widetilde{\cS}^{N_{\rmR}}}\left(\bz^{(\ell)} - \frac{1}{\beta_{(\ell)}}\nabla_{\bphi}F_{\lambda}(\bz^{(\ell)} \mid \bphi^{(\ell)})\right), \\
\bz^{(\ell)} & = \bphi^{(\ell)} + \alpha_{(\ell)}(\bphi^{(\ell)} - \bphi^{(\ell-1)}),
\end{align}
where the extrapolation sequence $\left\{\alpha_{(\ell)}\right\}_{\ell\geq 0}$ is the same as that in \eqref{eq:alpha}. Moreover, $\beta_{(\ell)}$ should satisfy a similar descent condition in \eqref{eq:descent property} given by
\begin{align}\label{eq:beta}
& F_{\lambda}(\bphi^{(\ell+1)} \mid \bphi^{(\ell)}) \leq F_{\lambda}(\bz^{(\ell)} \mid \bphi^{(\ell)})  + \frac{\beta_{(\ell)}}{2}\left\|\bphi^{(\ell+1)}-\bz^{(\ell)}\right\|^2 \ntb
& \quad\quad\quad\quad\quad + \left\langle \nabla_{\bphi}F_{\lambda}(\bz^{(\ell)} \mid \bphi^{(\ell)}), \bphi^{(\ell+1)}-\bz^{(\ell)}\right\rangle,
\end{align}
and then $\beta_{(\ell)}$ can be found via backtracking line search. Notice that in the GEMM method, although the MM subproblems are addressed inexactly, the same stationary convergence as the traditional exact MM method can also be guaranteed \cite{li2016sum}. In addition, with the number of APG iterations being limited to one, the GEMM method converges much faster compared with the exact MM method (executed via APG) \cite{shao2019a}, as will be demonstrated in \secref{sec:numerical}.

To facilitate the implementation of the GEMM method, more details about the projection operation $\Pi_{\widetilde{\cS}^{N_{\rmR}}}(\bphi)$ and the negative gradient $\nabla_{\bphi}F_{\lambda}$ in \eqref{eq:GEMM} will be further explained in the following. Firstly, the projection $\Pi_{\widetilde{\cS}^{N_{\rmR}}}(\bphi)$ onto $\widetilde{\cS}$ is performed in an element-wise manner. Therefore, it is sufficient to only consider the simpler projection $\Pi_{\widetilde{\cS}}(\bphi)$. In particular, for both $\cS_1$ and $\cS_2$, the nice geometric structure of $\widetilde{\cS}$ enables the closed-form expression of $\Pi_{\widetilde{\cS}}(\bphi)$. More specifically, for the CPS case, $\Pi_{\widetilde{\cS}_1}(\phi)$ is given by
\begin{align}\label{eq:projection1}
\Pi_{\widetilde{\cS}_1}(\phi) = \left\{ \begin{array}{l}
\phi ,\quad\quad\quad \left| \phi \right| \leq 1 \\
\phi / \left| \phi \right|, \quad \left| \phi \right| > 1.
\end{array} \right.
\end{align}
Meanwhile, for the DPS case, $\Pi_{\widetilde{\cS}_2}(\phi)$ is computed as
\begin{align}\label{eq:projection2}
\Pi_{\widetilde{\cS}_2}(\phi) = \mathrm{e}^{\jmath\frac{2\pi n}{\tau}} \left(\left[ \Real{\widetilde{\phi}}\right]^{\cos(\pi / \tau)}_{0} +
\jmath \left[ \Imag{\widetilde{\phi}}\right]^{\sin(\pi / \tau)}_{- \sin(\pi / \tau)} \right),
\end{align}
where $n = \left\lfloor \frac{\angle\phi+\pi / \tau}{2\pi / \tau} \right\rfloor$, $\widetilde{\phi} = \phi \mathrm{e}^{-\jmath\frac{2\pi n}{\tau}}$, and $[x]_a^b \triangleq \min\left\{b,\max\left\{x,a\right\}\right\}$ is the thresholding operator. Secondly, the gradient $\nabla_\bx f(\bx)$ follows the standard definition if $\bx \in \R^n$, and is defined as $\nabla_\bx f(\bx) = \nabla_{\Real{\bx}} f(\bx) + \jmath \nabla_{\Imag{\bx}} f(\bx)$ if $\bx \in \C^n$. Hence, the gradient $\nabla_{\bphi}F_{\lambda}$ is accordingly derived as follows:
\begin{align}\label{eq:gradient}
\nabla_{\bphi}F_{\lambda}(\bz^{(\ell)} \mid \bphi^{(\ell)})= 2 \left( \bB \odot \bA^T \right)\bz^{(\ell)} - 2\bc^* - 2 \lambda \bphi^{(\ell)}.
\end{align}

Lastly, the whole procedure of the NSP-based GEMM method to handle $\cP_{6a}$ is summarized in \alref{alg:GEMM}. Empirically, it is preferable to initialize the penalty parameter $\lambda$ to a relatively small value such that an ill-posed problem can be circumvented \cite{shao2019a}. As the rounds of GEMM iterations increase or when the distance between two consecutive iterations is smaller than the threshold, we then gradually increase $\lambda$ until it is large enough to meet the equivalence condition stated in \propref{prop:prop3}. Note that the convergence guarantee of \alref{alg:GEMM} can be proved using a similar approach as in \cite{shao2019a}.

\begin{algorithm}[h]
\caption{NSP-based GEMM Method for $\cP_{6a}$.}
\label{alg:GEMM}
\begin{algorithmic}[1]
\Require Integers $J \geq 1$, $c>1$, an initial penalty parameter $\lambda > 0$ and its breakpoint $\lambda_{\mathrm{upp}} > 0$, an extrapolation
sequence $\left\{\alpha_{(\ell)}\right\}_{\ell\geq 0}$, feasible phase shift vector $\bphi^{(\ell)} \in \widetilde{\cS}^{N_{\rmR}}$, iterative index $\ell = 0$, and threshold $\varepsilon$.
\State Initialize $\bphi^{(-1)} = \bphi^{(0)}$.
\Repeat
\For{$i=1$ to $J$}
\State Update $\bz^{(\ell)} = \bphi^{(\ell)} + \alpha_{(\ell)}(\bphi^{(\ell)} - \bphi^{(\ell-1)})$.
\State Decide $\beta_{(\ell)}$ via backtracking line search.
\State Update $\bphi^{(\ell+1)} = \Pi_{\widetilde{\cS}^{N_{\rmR}}}\left(\bz^{(\ell)} - \frac{1}{\beta_{(\ell)}}\nabla_{\bphi}F_{\lambda}(\bz^{(\ell)} \mid \bphi^{(\ell)})\right)$.
\If{$\left\|\bphi^{(\ell+1)} - \bphi^{(\ell)} \right\| < \varepsilon$}
\State $\lambda = \lambda c$.
\EndIf
\State Set $\ell = \ell + 1$.
\EndFor
\State $\lambda = \lambda c$.
\Until{$\lambda \geq \lambda_{\mathrm{upp}}$.}
\Ensure The solution $\bphi^* = \bphi^{(\ell)}$ to $\cP_{6a}$.
\end{algorithmic}
\end{algorithm}

\subsection{Overall Algorithm and Complexity Analysis}\label{sec:overallopt}
Combining the proposed methods for finding the solutions of $\bQ$ and $\bPhi$, which are respectively described in the above Sections \ref{sec:Optimization_Q} and \ref{sec:Optimization_Phi}, we reach a complete RE maximization approach for RIS-aided multi-user MIMO uplink transmissions with partial CSI and summarize the approach in \alref{alg:RE_maximization}.

\begin{algorithm}[h]
\caption{AO-based RE Maximization Method.}
\label{alg:RE_maximization}
\begin{algorithmic}[1]
\Require Feasible $\bLambda^{(0)}$, $\bPhi^{(0)}$, iterative index $t=0$.
\Repeat
\State Update $\bQ$ with given $\bPhi^{(t)}$:
\State \ \ Set $\bV_{k}^{(t+1)} = \bV_{2,k}$, $\forall k$, according to \propref{prop:beam_domain_optimal}.
\State \ \ Calculate the asymptotic system SE, $\SEde \left( \bLambda^{(t)} \right)$, and the auxiliary parameters, $(\bgamma,\bpsi)$, with $\bLambda^{(t)}$ and $\bPhi^{(t)}$ by \alref{alg:Deterministic}.
\State \ \ Update $\bLambda^{(t+1)}$ via solving $\overline{\cP}_{\bLambda}$ by \alref{alg:Quadratic}.
\State \ \ Update $\bQ_k^{(t+1)} = \bV_{2,k}^H  \bLambda_k^{(t+1)} \bV_{2,k}, \ \forall k \in \cK$.
\State Update $\bPhi$ with given $\bQ^{(t+1)}$:
\State \ \ Update the DE auxiliary parameter, $\bpsi$, with $\bPhi^{(t)}$ and $\bLambda^{(t+1)}$ by \alref{alg:Deterministic}.
\State \ \ Update $\bPhi^{(t+1)}$ via addressing $\cP_5$ by \alref{alg:BCD} and \alref{alg:GEMM}.
\State Set $t = t + 1$.
\Until{Some stopping criterion is satisfied.}
\Ensure Transmit covariance matrices $\bQ=\bQ^{(t)}$ and the RIS phase shift matrix $\bPhi=\bPhi^{(t)}$.
\end{algorithmic}
\end{algorithm}

It is worth noting that by setting different values for the weight $\beta$ in \alref{alg:RE_maximization}, we can attain different transmission schemes for EE-SE tradeoff. In addition, \alref{alg:RE_maximization} can be specialized into the approaches that maximize the system EE or SE, although it is initially designed for RE maximization. To be specific, by setting $\beta = 0$, \alref{alg:RE_maximization} can be straightforwardly reduced to EE maximization (with bandwidth normalization). Moreover, if the denominator of EE becomes a constant, maximizing RE is equivalent to maximizing SE. Therefore, by setting $\xi_k = 0$, $\forall k$, \alref{alg:RE_maximization} can tackle the special case that maximizes SE. Such modification has some impact on the optimization of power allocation strategies performed by \alref{alg:Quadratic}. Specifically, since $\overline{\cP}_3$ is non-fractional and convex, the optimal solution can be attained by \alref{alg:Quadratic} within just a single iteration.

Now, we discuss the complexity of the proposed algorithms as follows.
The main structure of \alref{alg:RE_maximization} is based on the AO method, which requires a total of $I_{\mathrm{AO}}$ iterations. In addition, due to the fast convergence rate of the DE method \cite{Couillet11Random} in \alref{alg:Deterministic}, the per-iteration complexity of \alref{alg:RE_maximization} is mainly composed by \alref{alg:Quadratic} for optimizing $\bLambda$ and \alref{alg:BCD} for optimizing $\bPhi$.
For \alref{alg:Quadratic}, we assume a maximum total of $I_{\mathrm{QT}}$ iterations in the quadratic transformation method where each iteration needs to tackle a convex program with $N$ variables. Hence, the complexity of \alref{alg:Quadratic} can be asymptotically estimated as $\cO(I_{\mathrm{QT}} N^p)$, where $1 \le p \le 4$ for standard convex program solutions \cite{huang2019reconfigurable}.
As for \alref{alg:BCD}, we assume that the BCD method requires to perform $I_{\mathrm{BCD}}$ iterations, each comprises three major optimizations in terms of $\bW_{\mathrm{h}}$, $\bU_{\mathrm{h}}$, and $\bPhi$, respectively. Note that the optimal results of $\bW_{\mathrm{h}}$ and $\bU_{\mathrm{h}}$ can be obtained in closed-form using \eqref{eq:optimal_W} and \eqref{eq:optimal_U}, respectively, and the corresponding complexity of calculating $\bW_{\mathrm{h}}^{\mathrm{opt}}$ and $\bU_{\mathrm{h}}^{\mathrm{opt}}$ are given by
$\cO(M^3)$ and $\cO(N_{\rmR}^3)$, respectively.
Then, we focus on the complexity of the NSP-based GEMM method in \alref{alg:GEMM} for optimizing $\bPhi$. The main contributor of the complexity at each iteration relies on the computation of the gradient in \eqref{eq:gradient}, which is approximately ${\mathcal{O}}(N_{\mathrm{R}}^2)$ \cite{shao2019a}. Thus, assuming $N_{\mathrm{R}}> M$, the complexity of optimizing $\bPhi$ in each iteration of the BCD method is $ {\mathcal{O}}(N_{\mathrm{R}}^3+I_{\mathrm{GEMM}}N_{\mathrm{R}}^2)$ where $I_{\mathrm{GEMM}}$ is the total number of iterations in \alref{alg:GEMM}.
In conclusion, the overall complexity of the AO-based \alref{alg:RE_maximization} is estimated as
$\cO\left( I_{\mathrm{AO}}\left(I_{\mathrm{QT}} N^p + I_{\mathrm{BCD}}(N_{\mathrm{R}}^3+I_{\mathrm{GEMM}}N_{\rmR}^2)\right)\right)$.

\begin{remark}
Notice that although \alref{alg:RE_maximization} is tailor-made for the scenario with partial CSI, it can also be extended to the scenario with instantaneous CSI after slight modification. Specifically, if instantaneous CSI is available, no expectation operations are required in the instantaneous SE and EE expressions in \eqref{eq:ergodic_SE} and \eqref{eq:ergodic_EE}, respectively. Hence, via replacing the DE-based asymptotic objective functions with the corresponding instantaneous ones, \alref{alg:RE_maximization} can handle the RE maximization problem for the case of instantaneous CSI. In addition, although we focus on the channel model without a direct UT-to-BS link, the proposed \alref{alg:RE_maximization} can still be applied to the more general channel model considering both RIS-assisted and direct links. In this case, although it is not easy to obtain the UTs' transmit directions in closed-form, one can modify \alref{alg:Quadratic} slightly to tailor for the optimization of the transmit covariance matrices.
\end{remark}

\section{Numerical Results}\label{sec:numerical}
In this section, simulation results are provided to verify the performance of the proposed optimization framework for the considered RIS-aided multi-user MIMO uplink communication. As for the small scale fading of the RIS-to-BS and UT-to-RIS channels, the suburban macro scenario is considered and we set the primary statistical channel parameters utilizing the 3GPP spatial channel model \cite{Salo05MATLAB}. Then the channel statistics, $\bOmega_{k}$, can be obtained using some existing methods, e.g., \cite{Gao09Statistical}. Meanwhile, as regards the large scale fading, we assume the path loss of $-120$ dB for all the end-to-end composite UT-RIS-BS channels, i.e., $\bH_{1} \bH_{2,k}$, $\forall k$. Unless further specified, we list the major simulation parameters in \tabref{tab:simulation_set_up} \cite{huang2019reconfigurable,Bjornson15Optimal}. Without loss of generality, the individual transmit power budgets for all UTs are assumed identical in the simulations, i.e., $P_{\max,k} = P_{\max}$, $\forall k$.

\newcolumntype{L}{>{\hspace*{-\tabcolsep}}l}
 \newcolumntype{R}{c<{\hspace*{-\tabcolsep}}}
 \definecolor{lightblue}{rgb}{0.93,0.95,1.0}
 \begin{table}[!t]
  \caption{Simulation Setup Parameters}\label{tab:simulation_set_up}
  \centering
\footnotesize
  \ra{1.2}
 \begin{tabular}{LR}
  \toprule
  Parameters &   {Values}\\
  \midrule
  \rowcolor{lightblue}
  Number of UTs $K$ & {$4$} \\
  Number of UT antennas $N_k$, $\forall k$ & {$2$} \\
  \rowcolor{lightblue}
  Number of RIS reflecting units $N_{\rmR}$ & {$32$} \\
  Number of BS antennas $M$ & {$8$} \\
  \rowcolor{lightblue}
  System bandwidth $W$ & {$10$ MHz} \\
  Background noise variance at the BS $\sigma^2$ & {$-96$ dBm} \\
  \rowcolor{lightblue}
  Amplifier inefficiency factor $\frac{1}{\xi_k}$, $\forall k$ & {$1/0.3$} \\
  Static power consumption of each UT $P_{\mathrm{c},k}$, $\forall k$ & {$10$ dBm} \\
  \rowcolor{lightblue}
  Hardware dissipated power at the BS $P_{\mathrm{BS}}$, $\forall k$ & {$39$ dBm} \\
  Per-element static power at the RIS $\Ps(1),\Ps(2),\Ps(+ \infty )$ & {$5$, $15$, $25$ dBm} \\
  \rowcolor{lightblue}
  Accuracy setting  $\varepsilon$ & {$10^{-4}$} \\
  \bottomrule
 \end{tabular}
 \end{table}

\subsection{Convergence Performances}
Fig. 2(a) illustrates the average convergence performance of the AO-based optimization framework in \alref{alg:RE_maximization}. Furthermore, the average convergence behaviors of the two major algorithms invoked in \alref{alg:RE_maximization}, including \alref{alg:Quadratic} for optimizing $\bLambda$ and \alref{alg:BCD} for optimizing $\bPhi$, are illustrated in Figs. 2(b) and 2(c), respectively. The results demonstrate that in typical power budget regions, all these algorithms enjoy fast convergence rates. In particular, \alref{alg:Quadratic} usually converges after only one step, so does \alref{alg:BCD} for the cases of small $P_{\max}$.

\begin{figure}[!t]
\centering
\subfloat[]{\centering\includegraphics[width=0.45\textwidth]{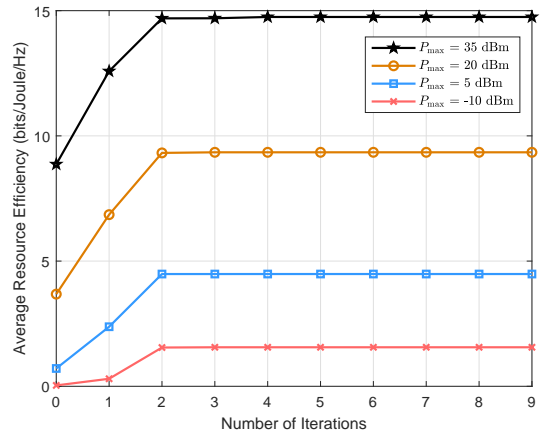}
\label{fig:convergence_AO}}
\hfill
\subfloat[]{\centering\includegraphics[width=0.45\textwidth]{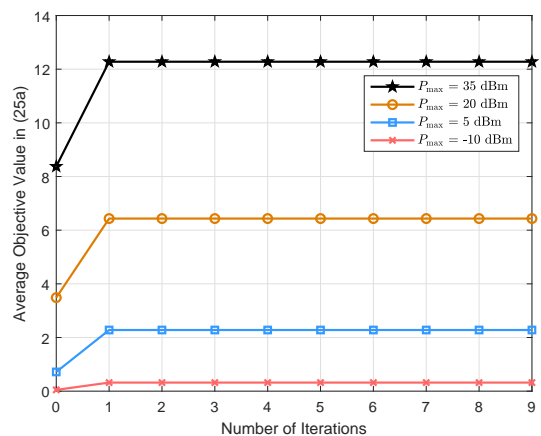}
\label{fig:convergence_Quadratic}}
\hfill
\subfloat[]{\centering\includegraphics[width=0.45\textwidth]{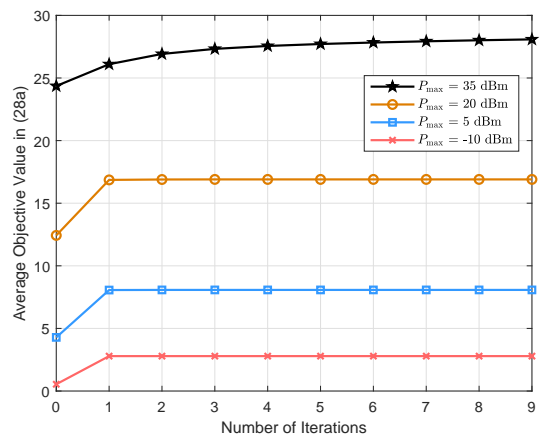}
\label{fig:convergence_MSE}}
\caption{Average convergence performances versus the number of iterations for the CPS case ($\beta / \Ptot = 0.5$). (a) Convergence of the AO-based RE maximization method in \alref{alg:RE_maximization}; (b) Convergence of the quadratic transformation in \alref{alg:Quadratic}; (c) Convergence of the iterative WMMSE method in \alref{alg:BCD}.}
\label{fig:convergence}
\end{figure}

\subsection{SE Maximization Design}
\begin{figure}
\centering
\includegraphics[width=0.45\textwidth]{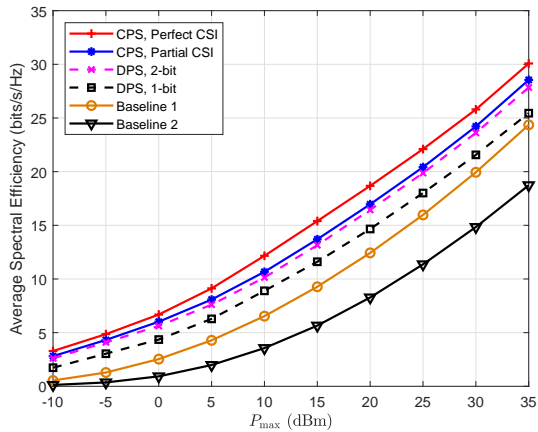}
\caption{Average SE performance versus $\Pmax$ achieved by SE maximization algorithm.}
\label{fig:SE_SEmax}
\end{figure}

To evaluate the effectiveness of the proposed framework in \alref{alg:RE_maximization}, we firstly apply it to SE maximization design via setting $\xi_k = 0$, $\forall k$, which is a special case of \alref{alg:RE_maximization} as described in \secref{sec:overallopt}. In \figref{fig:SE_SEmax}, we present the average SE performance versus $P_{\max}$ under different resolutions of RIS phase shifters. In addition, the transmission scheme with full instantaneous CSI is also considered and serves as the benchmark. In order to show the necessity of the joint power allocation and phase shift adjustment for SE enhancement, we also present the SE results of two baseline schemes. For the case of baseline 1, the RIS phase shift matrix is fixed as an identity matrix, i.e., $\bPhi = \bI_{N_\rmR}$. For this baseline scheme, since $\bPhi$ is fixed, only power allocation remains necessary and is implemented by \alref{alg:Quadratic}. For the case of baseline 2, except for setting $\bPhi$ to be $\bI_{N_\rmR}$, an equal power allocation scheme (with full power budgets) is also considered. Unsurprisingly, adopting an optimized $\bLambda$ brings about remarkable SE gains, and so does an optimized $\bPhi$, which happens even for the DPS case with the lowest bit resolution. Moreover, it is observed that the SE performance achieved by SE maximization design increases with the resolution $b$. As expected, the infinite resolution case attains the most SE gain. However, the DPS case with ``2-bit'' RIS phase shifters performs very close to the CPS case but needs much lower energy consumption, which will be demonstrated in the next subsection.

\subsection{EE Maximization Design}

\begin{figure}
\centering
\includegraphics[width=0.45\textwidth]{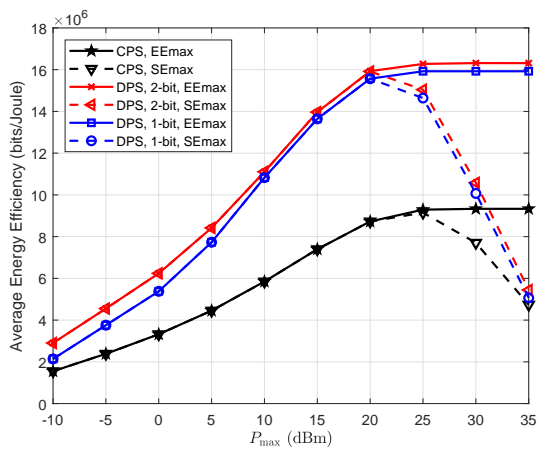}
\caption{Average EE performance versus $\Pmax$ achieved by EE maximization algorithm.}
\label{fig:EE_EEmax}
\end{figure}

We consider the EE maximization design for the RIS-aided system by means of utilizing \alref{alg:RE_maximization} with the weighting factor set as $\beta = 0$. The EE performance of the EE maximization design is reported versus $P_{\max}$ in \figref{fig:EE_EEmax}, where we also provide the results achieved by the SE maximization design for comparison. For the EE maximization design, the results exhibit that EE increases with $P_{\max}$ only when $P_{\max}$ is smaller than a threshold and then EE approaches a constant when $P_{\max}$ is larger than the threshold value, which is true for both CPS and DPS cases with different resolutions. This behavior is explained by the reason that there exists a unique maximizer of the transmit power for EE maximization and the system EE saturates when the power budget exceeds the value of this maximizer. Consequently, the actual transmit power remains constant when the maximum EE is attained, leading to saturated EE and SE in high $P_{\max}$ regimes. For the SE maximization design, the corresponding EE also increases with $P_{\max}$ when $P_{\max}$ is small but eventually declines rapidly when $P_{\max}$ is large since all the available power budget is exhausted to maximize SE so that the exceedingly large transmit power leads to a rapid drop in EE. Moreover, it is worth noting that EE does not always increase with the RIS phase shifter resolution $b$. Although applying the RIS with higher resolution phase shifters could attain higher SE performance, as shown in \figref{fig:SE_SEmax}, it also results in a substantially higher static hardware-dissipated power bringing negative influence on EE. Hence, applying a RIS with DPSs might be more energy efficient than the one with CPSs.

\subsection{EE-SE Tradeoff}
\begin{figure}
\centering
\includegraphics[width=0.45\textwidth]{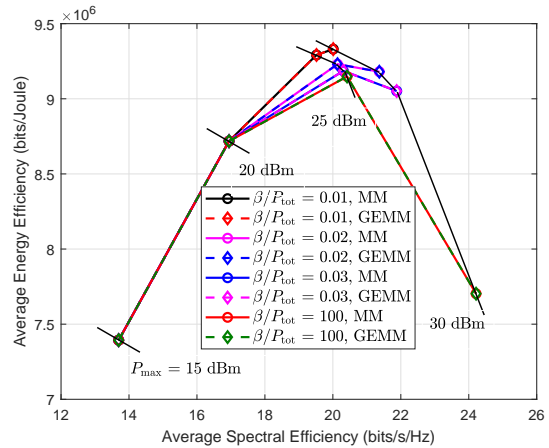}
\caption{Average EE-SE tradeoffs with different weighting factors for the CPS case under different transmit power budgets.}
\label{fig:tradeoff}
\end{figure}

The EE-SE tradeoffs attained by \alref{alg:RE_maximization} under different power budgets are demonstrated in \figref{fig:tradeoff}. In addition, different values of $\beta$ are considered to investigate the impact of the weighting factor on both EE and SE. It is observed that the EE-SE tradeoffs under different weighting factors are nearly the same when $P_{\max}$ is smaller than $25$ dBm. As can be seen in the above subsection, the results of EE and SE maximization designs are nearly identical when $P_{\max}$ is low because both EE and SE are maximized when all available power is exhausted. Hence, varying $\beta$ has no observable impact on the EE-SE tradeoff for low power budgets. In contrast, when $P_{\max}$ is high, the proposed RE maximization approach provides different tradeoffs to balance between EE and SE, hinging on the value of $\beta$. Specifically, increasing $\beta$ leads to a higher EE but a smaller SE while decreasing $\beta$ results in a smaller EE but a larger SE. This is because more focus is devoted to SE for the case of a larger $\beta$, and thus more power budget is utilized to increase SE. In particular, for the two extreme cases of $\beta / P_{\tot} = 0.01$ and $\beta / P_{\tot} = 100$, \alref{alg:RE_maximization} essentially performs solely the EE and SE maximization, respectively.

\newcolumntype{L}{>{\hspace*{-\tabcolsep}}l}
\newcolumntype{R}{c<{\hspace*{-\tabcolsep}}}
\begin{table}[!t]
\caption{Average Running Time (in sec) for the Update of $\bPhi$} \label{tab:GEMM_vs_MM}
\centering
\footnotesize
\ra{1.3}
\begin{center}
%\begin{tabular}{|m{2cm}|m{2cm}|m{2cm}|m{2cm}|}
\begin{tabular}{LccR}
\toprule
{$N_{\rmR}$ \quad} & {\qquad 16 \qquad} & {\qquad 32 \qquad} & {\qquad 64} \\
\midrule
{MM \quad} & {\qquad 0.48 \qquad} & {\qquad 0.79 \qquad} & {\qquad 1.30} \\
{GEMM \quad} & {\qquad 0.26 \qquad} & {\qquad 0.35 \qquad} & {\qquad 0.58} \\
\bottomrule
\end{tabular}
\end{center}
\end{table}

For comparison, we also provide the results obtained by the exact MM method, in which each MM subproblem is exactly solved by multiple APG iterations. The performances of the GEMM and the exact MM methods are almost identical, as shown in \figref{fig:tradeoff}. Moreover, \tabref{tab:GEMM_vs_MM} provides the runtime of these two methods. Here, we set $K=4$, $N_k=2$, $\forall k$, $M=32$, and $\beta / P_{\tot} = 0.01$. We conduct the simulations via the MATLAB 2018a on a desktop computer featuring 3.2 GHz Intel i7-8700 processor with 16 GB RAM. We can observe that GEMM runs much faster than MM. More specifically, while enjoying similar performance, the runtime of the iteration-limited GEMM method is nearly half of that of the exact MM method.

\section{Conclusion}\label{sec:conclusion}
In this paper, we investigated the transmission scheme for EE-SE tradeoff in the RIS-aided multi-user MIMO uplink system with the consideration of partial CSI. Specifically, to achieve an EE-SE tradeoff, we studied the precoding design at the UT sides and the reflecting phase shift adjustment to maximize the system RE, under both continuous- and discrete-phase shifts at the RIS. To handle the design optimization problem, we developed a sequential optimization framework by leveraging the AO method. For the precoding design with fixed RIS phase shifters, we first identified the optimal transmit subspaces at the UT sides with closed-form solutions and then found the asymptotically optimal power allocation strategies based on an asymptotic SE expression. For the RIS phase shift adjustment with fixed input covariance matrices, we proposed an iterative MMSE method combining with an inexact APG-based GEMM method, which is applicable to both CPS and DPS cases. Numerical results confirmed the efficiency of the developed optimization framework for the RE maximization (as well as the EE or SE maximization). In particular, the proposed scheme significantly increased SE compared to those employing equal power allocation or fixed RIS phase shifters. Moreover, substantial energy saving and outstanding EE gains were achieved by utilizing the RIS with DPSs compared to the case with CPSs.


\end{document}